\documentclass[aps,prb,showpacs,twocolumn,groupedaddress,floatfix]{revtex4}

\usepackage{graphics}
\begin{document}

%\preprint{}

\title{ Specific Heat of Liquid Helium in Zero Gravity
very near the Lambda Point }


\author{J. A. Lipa}
\email[]{jlipa@stanford.edu}
\author{J. A. Nissen}
\author{D. A. Stricker}
\affiliation{Physics Department, Stanford University, Stanford, CA}
\author{ D. R. Swanson }
\affiliation{Callida Genomics, Sunnyvale, CA}
\author{ T. C. P. Chui }
\affiliation{ Jet Propulsion Laboratory, Caltech, Pasadena, CA}

\date{\today}

\begin{abstract}
We report the details and revised analysis of an experiment 
to measure the specific heat
of helium with subnanokelvin temperature resolution near
the lambda point. 
The measurements were made at the vapor pressure spanning the region
from 22 mK below the superfluid transition to 4 $\mu$K above.
The experiment was performed
in earth orbit to reduce the rounding of the transition caused by
gravitationally induced pressure gradients on earth. 
Specific heat measurements were made deep in the asymptotic region to
within 2 nK of the transition.  No evidence of rounding
was found to this resolution. The optimum value of the critical
exponent describing the specific heat singularity
was found to be $\alpha = -0.0127\pm0.0003$. This is bracketed by
two recent estimates based on renormalization
group techniques, but is slightly outside the range of the error
of the most recent result.  The ratio of the
coefficients of the leading order singularity on the two sides of
the transition is A$^{+}$/A$^{-}$$ = 1.053\pm0.002$, 
which agrees well with a recent estimate. By combining the specific
heat and superfluid density exponents a test
of the Josephson scaling relation can be made. Excellent agreement
is found based on high precision measurements
of the superfluid density made elsewhere.  These results represent
the most precise tests of theoretical predictions
for critical phenomena to date.
\end{abstract}

%\keywords{}

\maketitle


\section{\label{sec:level1}Introduction \protect\\}


%Define special character (circle) from Kostelecky's Paper which
% is used for an Earth-base experiment. 
%These go with calligraphic A,B
\def\om{\omega}
\def\fr#1#2{{{#1} \over {#2}}}
%+++++++++++++++++++++++++++++++++++++++

The renormalization group (RG) theory \cite{Wilson1971} has long been accepted as the basis of our
understanding of critical phenomena due to its ability to deal with the problem of 
fluctuations on a wide range of length scales and to realistically
predict many quantities of experimental interest. However, it is well known that precise 
quantitative predictions with correspondingly accurate experimental tests are
few. Essentially all tests performed near gas-liquid critical points are unable to
give detailed attention to the asymptotic region close to the phase transition, where
the RG predictions are simplified and best established. Severe difficulties are
encountered here due to transition
broadening associated with gravity and
relaxation phenomena, limiting the useful temperature range for tests of the
theory. For example, near the critical point of a 1 mm high
sample of xenon, density gradients cause substantial distortion of the singularity for reduced
temperatures, $|t|$, $\lesssim10^{-4}$, where $t = 1-\text{T}/\text{T}_{c}$ 
and T$_{c}$ is the transition temperature.\cite{Moldover1979}
This results in the observation of effective
exponents which are perturbed from their asymptotic
values by higher order terms, providing only weak support for the theory.
To avoid this problem some experiments have been performed in space.
For example, recent measurements \cite{HauptStraub1999}
of the specific heat of SF$_{6}$ have extended about an order of magnitude
closer to the transition than ground based experiments. \cite{Edwardsthesis1984}
Relaxation phenomena now appear to inhibit further gains.
In contrast, at the lambda point of helium, the transition between
He I and He II, a much wider range of $|t|$ 
is accessible. On earth values
of $|t| \lesssim10^{-7}$ can typically 
be reached \cite{LipaPhyRevLett1983} before 
finite size effects become the limiting
factor. This has resulted in the lambda transition becoming the premier testing ground
for the RG theory of second order transitions. In space, the lambda transition is expected
to be sharp to $|t| \approx 10^{-12}$ in ideal
conditions. \cite{Lipa75thJubilee1983} Here
the challenge is more to develop a measurement technique than to obtain a suitable
sample.

Recently, improved predictions have been
derived for a number of universal quantities of three 
dimensional critical systems. \cite{Kleinert2000,Campostrini2001,StrosserDohm2000}
These predictions now approach the accuracy of the corresponding experimental
observations near the lambda point. This advance has uncovered a small but not entirely
negligible discrepancy between the heat capacity exponent, $\alpha$, measured in a space
experiment conducted in 1992 \cite{LipaPhyRevLett1996,LipaPhyRevLett2000} and 
the most recent predictions. Since this comparison
currently represents one of the core tests of the RG theory, we have completed a more 
detailed analysis of the experiment with the aim of reducing systematic bias in the
results as much as possible. We have also included additional measurements that were rejected
earlier due to uncertainties in their calibration. The results of this analysis are presented
here, along with a detailed report on the significant aspects of the experiment. We now
find better agreement with the predictions than previously. 

%--------------------------------------------------------------------------
\begin{table*}
\caption{\label{tab:table1}Comparison of predicted and observed values of universal
quantities near the lambda point.}
\begin{ruledtabular}
\begin{tabular}{ccccc}
 Quantity&Predicted&Reference&Observed&Reference\\
\hline
$\alpha$ & $-0.01126\pm 0.0010$ & 7 &$\sim -0.022\pm 0.006$ & 15,16,17 \\
$\alpha$ & $-0.0146\pm 0.0008$ & 8 & $-0.0105\pm 0.00038$ & 10,11 \\
$\zeta$ & $0.67015\pm 0.00027$ & 8 & $0.6705\pm 0.0006$ & 18 \\
$3\zeta+\alpha-2$ & $0$(exact) & 20 & $-0.0012\pm 0.0027$ & 5,18 \\
$\Delta$ & $0.529\pm 0.009$ & 25 & $ 0.5\pm 0.1$ & 22 \\
$P$ & $4.433\pm 0.077$ & 9 & $4.2\pm 0.1$ & 23 \\
$a_{c}^{+}/a_{c}^{-}$ & $1.6\pm 1.0$ & 24 & $1.0\pm 0.3$ & 23 \\
$a_{c}^{-}/\alpha a_{\rho}$ & $3.4\pm 0.1$ & 24 & $4.1\pm 0.2$ & 23 \\
\end{tabular}
\end{ruledtabular}
\end{table*}
%--------------------------------------------------------------------------

While the experiment was in
principle very straightforward, the constraints of operation 
in space dictated a number of
compromises in the instrument design and the data collection procedures, which increased
the complexity and the noise level. Because of the extreme cost of such experiments,
it is unlikely that the measurements will be independently verified for quite some
time. We have therefore attempted to provide sufficient 
detail in the sections below to allow a
reasonably complete assessment of the strengths and weaknesses of the experiment.

The remainder of this 
section is devoted to a brief
summary of the status of RG testing near the lambda point
focusing on static exponents, and a discussion
of the intrinsic limits expected in the present experiment. 
In Sec. II we describe the apparatus and in Sec. III
we discuss the calibration and heat capacity measurements performed 
on the apparatus prior to the flight mission.  
The sequence of the flight measurements is described in 
Sec. IV and some post-flight measurements
are discussed in Sec. V.  In Sec. VI we present the analysis of the 
data and discuss its significance. 
We summarize in Sec. VII.

\subsection{\label{sec:level1a}Background} 
The lambda transition of helium is the primary example of the universality
class with a two-dimensional
order parameter in three spatial dimensions (n=2, D=3) and has a strong divergence of
the correlation length, leading to easily measurable critical effects.
In addition, the transition has a number of unique
properties which can aid investigations. For example, the transition occurs
at a line rather than at a point on
the phase diagram, simplifying
the experimental requirements compared to a liquid-vapor critical point. Also the
compressibility is only weakly divergent,
substantially reducing gravity effects. On the low temperature side of the transition
the liquid is in the superfluid state,
essentially free of temperature gradients. These features have enhanced our ability
to perform a number of very high resolution investigations of the transition
region.

Of primary interest for static phenomena near the lambda point 
are the behavior of the 
specific heat, $\text{C}_{p}$, and the superfluid density, $\rho_{s}$.
In a quantitative analysis of the temperature dependence of 
these parameters it is necessary to deal
with non-asymptotic representations because data are obtained a finite distance
from $\text{T}_{\lambda}$.
In this region the RG theory predicts that \cite{Wegner1972} 
\begin{equation}
\text{C}_{p} =\frac{\text{A}^{\pm}}{\alpha}|t|^{-\alpha}( 1 + a_{c}^\pm|t|^{\Delta}
+ b_{c}^\pm|t|^{2\Delta} + .... ) + \text{B}^{\pm} \label{1}
\end{equation}
where the + and - signs refer to T $>$ T$_{\lambda}$  and  T $<$ T$_{\lambda}$
respectively, and
\begin{equation}
\rho_{s} =\rho_{o}t^{\zeta}(1+ a_{\rho}t^{\Delta}+
b_{\rho}t^{2\Delta} + .... ) ~~~  \text{T}<\text{T}_{\lambda}  \label{2}
\end{equation} 
Within the RG scheme, the quantities A$^{\pm}$, B$^{\pm}$, and $\rho_{o}$ as well as the correction
amplitudes $a_{c}^{\pm}$, $b_{c}^{\pm}$, $a_{\rho}$, $b_{\rho}$ depend on pressure-dependent
parameters of the statistical distribution of the order parameter, whereas the critical exponents
$\alpha$, $\zeta$ and
the confluent singularity exponent $\Delta$ are independent of these parameters, i.e. universal.
Certain ratios of the non-universal amplitudes
are also predicted \cite{Privam1991} to be universal,
for example, $\text{A}^{+}/\text{A}^{-}$, $a_{c}^{+}/a_{c}^{-}$, $a_{c}^{-}/a_{\rho}$.
It is also useful to define the quantity 
$P\equiv(1-\text{A}^{+}/\text{A}^{-})/\alpha$ which is relatively stable when
$\alpha$ is small. \cite{BarmatzPRevB1975} 
Using advanced analytical and numerical techniques it has been
possible to derive estimates for many of the
universal quantities, a number of which are summarized in Table I. Recently the
bounds on the theoretical value of $\alpha$ have been
reduced substantially, allowing a higher quality test of the theory. Two recent predictions for this
quantity are listed in the table.

In the case of $\alpha$ the experimental situation is
somewhat complex because of the differing
conditions of various measurements. A number of experiments are in approximate agreement,
but indicate a possible
difference between the value of  $\alpha$ at the vapor pressure and that 
along the lambda lines as a function of pressure
and $^{3}$He concentration.
Mueller et al. \cite{MuellerPhyRevB1976} obtained $\alpha = -0.026 \pm 0.004$ 
from measurements of the isobaric
expansion coefficient as a function
of temperature and pressure. 
Gasparini and Moldover \cite{GaspariniMoldover1975} performed measurements of the specific heat
along the $^{3}$He - $^{4}$He lambda line at constant $^{3}$He concentration. 
These data were fitted to Eq.(1) by Gasparini and Gaeta \cite{GaspariniGaeta1978} 
assuming $b_{c}^{\pm} = 0$.
Excluding the measurements for pure $^{4}$He, they
obtained $\alpha = -0.025$, in very good agreement with the expansion coefficient result.
On the other hand,
their pure $^{4}$He results gave $\alpha = -0.016$. Some of the early ground measurements
from this program \cite{LipaPhyRevLett1983}
gave $\alpha = -0.0127\pm0.0026$ at the vapor pressure, 
and the previous
analysis of the present experiment \cite{LipaPhyRevLett1996,LipaPhyRevLett2000}
gave $\alpha = -0.0105 \pm 0.00038$.
While the differences
are not large in absolute terms,
the exponent results along the lambda lines do not appear to be entirely compatible
with the vapor pressure measurements.
In Table I we have indicated an approximate range of $\alpha$ obtained from other
groups and the previous result from the present experiment.

Recent values of $\zeta$ obtained from second sound data at the vapor
pressure are $0.6705\pm0.0006$ by
Goldner et al. \cite{GoldnerAhlers1992} and $0.67016\pm0.00008$ 
by Adriaans. \cite{AdriaansLipa1994}
In the latter case the error quoted included only
the statistical uncertainty from the velocity measurements. 
The results for $\zeta$ and $\alpha$ can be combined to test
the exact scaling prediction \cite{Ahlersbook1976}
$3\zeta + \alpha - 2 = 0$. 
Combining the results
of refs. 5 and 18 (for example) we
obtain $3\zeta + \alpha - 2 = -0.0012\pm0.0027$ at
the vapor pressure.
As yet, no high accuracy measurements of $\zeta$ along the lambda lines have been reported.
However, higher order
departures from a truncated version of Eq.(\ref{2}) have
been shown \cite{SchlomsDohmEuroLett1987} to be
in good agreement with RG predictions.

\subsection{\label{sec:level1b} Factors Limiting the Exponent Determination} 
To obtain the best information on the divergence of the
specific heat at a cooperative transition it is generally considered desirable to collect the most
accurate data over the widest possible range. Both the inner and outer limits of this range are set
by practical issues which either increase the noise of the results or lead to bias in the derived
exponent.
In the present experiment we have the potential for obtaining a
statistical uncertainty $\sigma_{\alpha}\sim 0.0002$ 
so we need to consider bias effects that could lead to systematic 
errors $\Delta\alpha\sim 10^{-4}$.
We explore these and related issues in this subsection.

From an inspection of Eq.(\ref{1}) it can be seen that
the value obtained for $\alpha$ is likely to be more
reliable the smaller the value of $t$ at which the curve fitting is done.
This is because the higher
order terms in the expression tend to zero in the limit of small $t$. However, technical difficulties
increase in this region both due to measurement techniques and sample imperfections.
It is easy to show that on earth the lambda transition will be
severely rounded over a temperature interval of about 1.3 $\mu$K per centimeter of
hydrostatic head in a sample,
due to the slope of the lambda line, ($\partial$P/$\partial$T)$_{\lambda}$. 
For a sample of constant cross-section and height h, the effect on the specific heat can be
visualized as a convolution of a gravity-induced
temperature `window' 
$\Delta\text{T}_{\lambda} = \rho\text{gh}(\partial\text{T}/\partial\text{P})_{\lambda}$ with
the gravity-free function in Eq.(1), where $\rho$ is the 
density of the fluid and g is the
acceleration due to gravity.  This approximation is valid 
because over the relatively
small hydrostatic pressure heads encountered in typical calorimeters the pressure dependence
of the specific heat can be neglected. The situation can also be understood by examining the P-T phase
diagram of helium sketched in Fig. 1. The heavy tilted lines represent the vapor pressure curve
and the lambda line. Also shown is a vertical bar representing the states of the fluid in
an isothermal sample cell with a small vapor space and a moderate vertical height. It can
be seen that some of the states are closer to the lambda line than others, even at constant
temperature.  As the sample is warmed through the transition it enters a two-phase region
which persists over a temperature range given by $\Delta\text{T}_{\lambda}$. 
This hydrostatic pressure effect has been observed in numerous 
experiments. \cite{LipaPhyRevLett1983,Ahlers1968Barmatz1968} 

Clearly, smaller sample heights 
lead to less rounding, but soon a limit is reached where the height is so small that finite size
effects begin to affect the results. \cite{LipaPhyRevLett2000}
Normally this would occur at the angstrom level, but near
the lambda point the correlation length, $\xi$, characterizing these effects diverges
as $\xi \approx \xi_{0}t^{-2/3}$, magnifying the effect enormously.  This behavior
is intrinsic to all cooperative transitions, being associated with the physics of
the transition process in an essential way.

%****************************************************************
\begin{figure}
\includegraphics{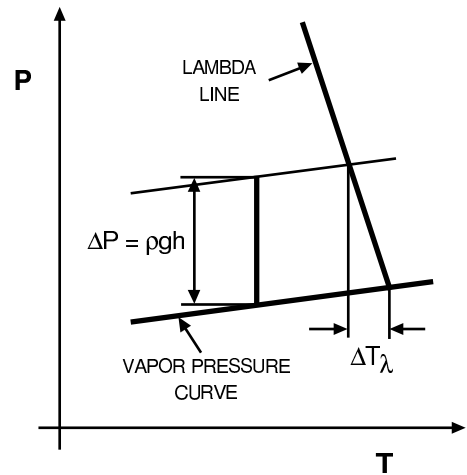}
\caption{\label{fig:epsart1}Phase diagram for liquid helium near the intersection of the vapor
pressure curve and the lambda line.}
\end{figure}
%****************************************************************

Using paramagnetic salt
thermometry techniques a temperature resolution of $<3\times10^{-10}\text{ K}/\sqrt{\text{Hz}}$ 
is now routinely available near the lambda point, and in many cases signal averaging
can improve this resolution substantially. 
To limit the hydrostatic rounding of the specific
heat at the lambda point to a 1 nK range, for example, the height of a helium sample would have
to be reduced to about $10^{-3}$cm.
But the divergence of $\xi$ at the transition implies that such a
thin sample would no longer exhibit
bulk behavior. 

By performing experiments in earth orbit the hydrostatic effect can be reduced substantially.
This allows the use of larger samples to reduce the finite size effect and thereby obtain useful
data at higher resolution. Unfortunately the acceleration environment on the shuttle is highly
variable so it is not possible to apply corrections for the hydrostatic effect.
A rough estimate of the magnitude of the effect can be obtained by considering the effect
of acceleration impulses of different duration. It is evident that impulses short compared to
the measurement time will have no effect on the result except for higher order dissipation effects.
On the other hand, impulses lasting as long as the measurement time will lead to a full hydrostatic effect.
Thus we are led to consider the low frequency portion of the acceleration spectrum as a
possible source of distortion of the heat capacity results. From the spectral measurements
made during the mission we found the acceleration amplitude to be $<3\times 10^{-5}$ g in the
region below 1 Hz. Using this as an upper limit and a sample size of 3.5 cm we obtain a transition
broadening $\Delta\text{T}_{\lambda} < 1.3\times 10^{-10}$ K. To determine the effect
on $\alpha$ we modeled the transition broadening as a gravity correction and obtained  
a perturbation $\Delta\alpha < 5\times10^{-5}$.
Both these effects are well below the level of detectability in the present experiment.

The first order departure from the bulk specific heat
in a finite system can be characterized as a surface specific heat term which has been measured
for one and two-dimensional geometries. Since the magnitude of this effect is proportional to the
surface area of a sample, the observed specific heat can be written approximately
as C$_{exp} =\text{C}_{p} + \text{AC}_{Surf}$, where A is the surface area and C$_{Surf}$ is
the surface specific heat per unit area.  The effect of the surface term on the value obtained
for $\alpha$ can easily be determined by modeling. For the calorimeter used here and
fitting over a range $10^{-9} < t < 10^{-2}$ we obtained a
perturbation $\Delta\alpha = 3.4\times10^{-5}$, using the finite size measurements 
of ref. 11 to estimate C$_{Surf}$.
Since this result was sensitive to the precision of the data close to the transition relative
to that far away, we used the actual experimental uncertainties as described in section VI
below to make the estimate. At the time the experiment was designed, the behavior
of the surface specific heat was not known with any certainty,
leading us to reduce the surface area relative to the sample
volume as much as possible.

As one includes data further from the transition the curve fitting procedure becomes more difficult
since more terms must be carried in the function to obtain an
accurate representation of the specific heat.
Neglect of these terms can lead to systematic bias in the remaining parameters that are evaluated.
First we consider the effect of neglecting a fourth order term
in Eq.(1), $c_{c}^{\pm}|t|^{3\Delta}$.
Assuming for estimating purposes that $c_{c}^{\pm} = 1$ and fitting our
data set with and without the term,
we find the bias from its neglect is $\Delta\alpha\sim 8\times10^{-5}$. 
While this can be ignored for now,
it shows the importance of an appropriate functional form, especially in high precision experiments.
Alternatively one can restrict the range
of the fit to reduce the
bias from the neglected terms, a
trade-off that sets the practical outer limit of the curve fitting region.

The experimental situation
is actually slightly more complex than indicated by Eq.(1) due to the
possibility of regular background terms
that have so far been ignored.  A more correct representation of the measured specific heat
is $C_{exp} = C_{p} + C_{reg}$ where 

%=============================================================
%\begin{equation}
\begin{eqnarray}
C_{reg} = c_{0} + c_{1}t + c_{2}t^{2} + ... \label{3}
\end{eqnarray}
%\end{equation}
%============================================================
in which  the $c_{i}$ are constants.  Clearly $c_{0}$ can be absorbed
into B in Eq.(1) but the effects of the other
terms need to be considered. From the wide range behavior of the specific
heat \cite{FairbankKellers1957} where fluctuation effects are
small we estimate $c_{1}\approx 2$ J/mole K. Thus for $t = 10^{-2}$ this
term contributes $\sim$ 0.07$\%$ to the
specific heat, a small but detectable amount. 
It is fortunate that the exponent of the third order
term in Eq.(1), 2$\Delta$ - $\alpha$ $\sim$ 1.07, is close to that of the 
second term in Eq.(3). Since the coefficients of the terms are fitted parameters, 
this allows us to consider the effect on $\alpha$ of
combining the two terms into one. To study this we made a set of simulated
data in which the term c$_{1}$ was
included in the generating function, but then ignored in the fitting procedure.
Again using the actual experimental uncertainties
we found a bias $\Delta\alpha\sim 5\times10^{-6}$
which is negligible here.  The value of $c_{2}$ is harder to determine, but it would appear to be
similar to $c_{1}$. Even at $t = 10^{-2}$ such a term contributes $\ll 0.01\%$ to the specific heat
and can be neglected. Thus for the present experiment a sufficiently good fitting function can be
obtained by absorbing $C_{reg}$ into the original expression in Eq.(1). The small price paid for doing
this is that the experimentally determined third order coefficients no longer represent quantities
of theoretical interest, and other coefficients may be slightly perturbed. In the analysis described
below we consider various fitting functions and ranges and evaluate their impact on the results.

A number of other factors can lead to bias in the exponent. For our experiment the most important
appear to be the calibrations of the temperature scale and the calorimeter heater circuit.
These issues are discussed in the relevant sections below. Other less significant issues
are the details of the thermodynamic path followed by the sample, the slight pressure
dependence of the parameters in Eq.(1), and gravity gradient effects.  

\section{\label{sec:level2}APPARATUS}

The apparatus developed for the experiment has been described elsewhere
in some detail. \cite{LipaCryogen1994}
In essence it consisted of a spherical copper calorimeter attached to a pair of high resolution
thermometers and enclosed in a thermal control system. The apparatus was located in a helium cryostat
rigidly mounted in the shuttle bay.
The central experimental issues for performing a high resolution specific heat experiment
are temperature resolution and thermal control. Conventional low temperature germanium
resistance thermometers (GRTs) are capable of
resolving to $\Delta t \sim 3\times 10^{-7}$ with
a power dissipation on the order of $10^{-7}$ W. Higher resolution would require increasing
the power input, which rapidly leads to unacceptable thermal offsets due to self-heating.
Near the lambda point, the measurement power can also cause variable temperature offsets
due to the strong temperature dependence of the thermal conductivity of the sample.
Since the goal of the experiment was to achieve a resolution of $\Delta t \sim 5\times 10^{-10}$ for
the specific heat measurements, it was essential to develop a new type of high resolution
thermometer (HRT). The device we constructed was based on paramagnetic salt thermometry
commonly used at very low temperatures and is described below.

The heat capacity measurements were made by measuring the temperature change
of the sample when a heat pulse is applied.  Typically, the energy applied
to the calorimeter was chosen so that the
temperature step was significantly smaller than
$\text{T} - \text{T}_{\lambda}$.
To make heat capacity measurements to $\sim 1\%$ accuracy near the
transition, it is necessary to control the energy input to 
the sample to $\sim 10^{-11}$ C Joules,
where C is the heat capacity of the sample.  In the present case, for a heater operating for
a few seconds, this corresponds to applied power uncertainties of $\sim 10^{-10}$ W.
In addition, during the drift period used to measure the corresponding temperature rise,
the uncontrolled fluctuations of the background power input need to be less than $\sim10^{-11}$ W on
time scales of $\sim 100$ sec. This shows that very careful thermal control of
the sample environment is necessary.  To achieve this, we built a four-stage
thermal control system which used HRTs on the inner stage as fine control sensors.
This control system was the major portion of the
low temperature apparatus that comprised the flight instrument.

A third item of great importance was a helium cryostat capable of operating in earth orbit.
We made use of a low temperature facility developed by NASA which can operate
near 1.7 K in zero gravity. \cite{LuchikElliottCryoEng1996}
The experiment was flown in late October 1992 on STS-52.
The instrument and electronics were built by Ball Aerospace based on prototypes
developed for earlier ground experiments. \cite{LipaPhyRevLett1983} 
The space-flight hardware was
constructed to meet rigorous design constraints including use of high reliability parts,
structural analyses to verify its ability to survive launch level vibrations, and
specialized construction techniques for operation in the thermal and vacuum conditions
of space. Particular attention was paid to shielding the instrument from electromagnetic
interference (EMI). Spurious EMI had the potential of heating the calorimeter in an
uncontrolled fashion and generating pick-up in the HRTs. Since it was not possible
to perform realistic EMI testing prior to the flight a conservative design approach
was taken where possible.

Careful attention was also given to the design and
fabrication of low temperature seals to minimize the possibility of vacuum leaks.
Sufficient low temperature sensors and heaters were provided that all critical
instrument operations had backup capability. This approach was reflected as far
as practical in the electronics with the exceptions of the computer, the communications
port and the power supply. Ball Aerospace also supplied a preliminary version of the
computer code used for controlling the experiment and transmitting real time data to
the ground during the mission. This software was designed to provide near-automatic
operation in case of communication problems during the mission and to accept a wide
range of commands to alter the parameters of the measurements.

\subsection{\label{sec:level2a}Thermal Control System}
The basic design of the thermal control system was similar to that used for earlier ground-based heat
capacity measurements. \cite{LipaPhyRevLett1983}
A number of structural changes were made to improve its utility and
its ability to survive launch. A schematic view of the system is shown in Fig. 3 of ref. 28. It consisted
of a vacuum can 20 cm in diameter and 60 cm long surrounding four thermal control stages in series
with the calorimeter (stage 5).  Four HRTs were housed in the lower part of the assembly and
surrounded by a thermal shield attached to the fourth stage of the thermal isolation system.
Two of the HRTs were attached to the calorimeter and the others to the thermal shield.
An end-corrected aluminum wire solenoid was wound on the outside of the lower portion of the
vacuum can to allow the application of a uniform magnetic field 
to the niobium flux tubes of the HRTs.

The primary structural element of the thermal isolation system was a tripod with legs of stainless
steel tubing attached to the lid of the vacuum can.  Three triangles of high thermal conductivity
copper intersected the legs of the tripod at intervals of about 3 cm.  These triangles formed the
first three stages of the isolation system and also served to stiffen the tripod.  They were
attached to the legs by brazing.  The ratio of the thickness of the 
copper triangles to the wall
thickness of the legs was chosen to provide good thermal grounding for heat flowing
down the legs.  A ratio of 10 was estimated to attenuate thermal leakage by
a factor of about $10^{3}$ at each joint.  The ratio of the thermal resistance
along a side of a triangle between two legs to that of a leg section between
triangles exceeded $10^{3}$, heavily attenuating any thermal gradients
that might exist in the tripod attachment assembly.

During the design of
the system it was found that the HRTs attached to the calorimeter might
be subjected to unacceptably high whiplash loads during launch. 
To alleviate this problem the structure between the vacuum can lid and
the tripod base was softened by the addition of a flexible plate. 
This had the side effect of making the instrument somewhat acceleration
sensitive at the resonant frequencies of the plate plus its load, primarily
near 55 Hz.  Vibration tests described elsewhere \cite{NissenLipaAIAA}
were undertaken to characterize
the heating due to variable accelerations and determine the effect of the resonances
in the structure during the mission.  The flexible plate was attached to
the vacuum can lid by six stainless steel rods. Due to the configuration of
the vacuum can and the cryostat, this lid was not in direct contact with the
helium bath, reducing the cooling power available to the thermal control system.
To rectify this problem we added flexible copper foil thermal links between the
tripod and a copper ring in the center of the cylindrical wall of the vacuum can
which was wetted by the helium. 

The innermost thermal control stage had a more complex
structure, consisting of a ring, shield and HRT support assembly formed from annealed
high thermal conductivity copper.  The shield completely surrounded the calorimeter and
the set of HRTs, acting as a baffle for stray thermal radiation and a shunt for
residual gas conduction. The calorimeter was located 
near the apex of a second stainless
steel tripod which was not in direct
contact with the legs of the first.

The temperatures of stages 1-3 were actively controlled with heaters and GRTs
configured in proportional-integral (PI) servo feedback loops. The tripods
and isolation stages form a thermal network.  With the temperature servos
operating, each GRT acts as a node or ground point in a low pass filter which
attenuates thermal variations from external sources such as the main helium bath.
The price paid for this active filtering is heat dissipation in the stages.
Thus as one moves inwards through the stages the uniformity of the thermal
environment transitions from being dominated by external effects to being
limited by internal dissipation.  By the time the third stage is reached,
the thermal inhomogeneities in the structure are expected to be dominated
by the power dissipated in the GRT and the heater.  Since the heater power
is typically 10-100 times that dissipated in a GRT this effect can be minimized
by reducing the offset between the operating temperatures of stages 2 and 3.
The temperatures of the stages are otherwise adjusted to provide sufficient
dynamic range for the servo systems to control transients.

The stage 4 thermal controller had two operating modes: a coarse mode with
a GRT as sensor and a fine mode with a HRT as sensor, both using a digital
PI control loop to drive a heater. In the fine operating mode it was found
that the long term stability of its temperature, T$_{4}$, was limited
primarily by drift of the HRT set point to about $5\times10^{-14}$ K/sec. 
Short term stability ($< 1000$ sec) was $\pm 3\times10^{-8}$ K, limited by the
ability of the servo system to reject the temperature fluctuations of stage 3.
We note that this level of control was much better in the apparatus built for
laboratory use, \cite{LipaPhyRevLett1983} with the 
deterioration traced to the stiffening of the
flight apparatus for launch survival. Nevertheless, the degree of thermal
control of the calorimeter was adequate. The effect of the fluctuations of T$_{4}$ on
the measurements can be estimated roughly as follows. The thermal resistance
between the calorimeter and stage 4 was found to be $2.18\times 10^{4}$ K/W.
Thus near the lambda point, where the sample heat capacity exceeded 50 J/K, the
time constant, $\tau$, for heat transfer between the calorimeter and stage 4
was $>$ 10$^{6}$ sec. The transient behavior of this part of the 
system can be modeled as
a low-pass thermal filter with a single pole roll-off. The amplitude of the
swings of the calorimeter temperature, T$_{5}$,
is given by
$\Delta\text{T}_{5} \approx \Delta\text{T}_{4}/2\pi\tau\text{f}_{4}$,
where f$_{4}$ is the frequency of the fluctuations of T$_{4}$. 
With f$_{4}\sim 10^{-2}$ Hz we
obtain $\Delta\text{T}_{5}\sim 5\times10^{-13}$ K. In ground testing it was found that with
fine adjustment of T$_{4}$, T$_{5}$ could be stabilized to a drift rate
of $< 10^{-12}$ K/sec for up to one hour. This was more than adequate to
achieve the desired level of uncertainty in the heat input during a heat
capacity measurement. To minimize thermal gradients in stage 4 the temperature
difference T$_{4}$ - T$_{3}$ was typically reduced to 1 mK, limited by the calibration
uncertainties over the operating range of the experiment
and the control requirements of the stage 4 servo.
A typical temperature profile of the thermal control system consisted of stage 5 at
a given operating temperature, stage 4 in thermal equilibrium with it, and
stages 1, 2 and 3 controlled at 30, 10 and 1 mK cooler than stage 4.

Thermal gradients in the calorimeter assembly were reduced by deactivating the
GRTs during high resolution measurements and using the heater only when step
temperature changes were desired.  With T$_{5}$ constant the potential sources
of residual temperature gradients are then the 
effects of stray electrical power pickup,
internal dissipation in the mechanical structure, gas conduction, and charged
particle heating in space. During ground operations the limiting factor on
thermal control of the calorimeter appeared to be variations in stray power
pickup in the circuits attached to it.  To reduce this effect all leads to the
calorimeter were thermally anchored at all stages using sapphire posts, which
minimized the capacitance to ground.  Also the leads were made of twisted pairs
and were equipped with capacitance shunts to ground where they entered the helium
cryostat, which formed a complete metallic enclosure around the instrument.
For the wiring to other stages, similar techniques were used, except that the
sapphire was replaced with copper.  The calorimeter was also thermally connected to
stage 4 by four capillary gas lines and two HRT pickup loops with niobium-titanium shields.
Special clamps were built to thermally anchor these lines to the isolation stages
by sandwiching the lines between indium foil for good thermal contact.  
The length of the clamped sections was at least twenty times
the diameter of the largest line.  All leads and lines were fabricated from low
thermal conductivity materials.  In the case of leads to heaters, superconducting
niobium-titanium wire was used, both to allow accurate heat input measurements,
and to avoid dissipation elsewhere in the thermal control system. With these
precautions, the stray power variations seen in ground testing were generally
below 10$^{-11}$ W over periods of hours.

\subsection{\label{sec:level2b}Calorimeter}
The helium sample was contained in a 3.5 cm diameter spherical cavity cut from
very high purity (6N) copper \cite{Nippon6N} by electrical discharge machining. This
technique was used to preserve the highly annealed state of the copper and
its associated high thermal conductivity at low temperatures. From the measured
electrical resistance ratio of $\sim 5700$ between $300$ and $4.2$ K we 
estimated \cite{HurstLankford1984} its thermal conductivity
at T$_{\lambda}$ to be $\sim 140$ W/cm K.  The cavity was formed in two pieces
that were electron-beam welded together. The inside surface was left somewhat
rough to promote thermal coupling with the helium.  A magnified view of this
surface is shown in Fig. 2. 

An approximately-to-scale cross-sectional view of the 
calorimeter is shown in Fig. 3.
The base of a valve assembly was electron-beam welded
to the top of the calorimeter. The valve was sealed by pressing a gold-coated
copper flat against a circular copper knife-edge. The sealing pressure was set by
adjusting the length of a stack of 16 Belleville washers
and the valve was opened by a pneumatic
actuator. A high degree of cleanliness was required in the knife-edge area to
obtain a reliable superfluid-tight seal. A filling capillary passed through the
thermal control system to a second valve on the outside of the cryostat. A charcoal
adsorption pump was attached to the capillary through a filter to trap any helium
leakage through the low temperature valve. The base of the calorimeter provided
attachment points for a pair of HRTs. A beryllium-copper support flange was welded around
the bottom perimeter of the calorimeter to provide attachment points for the support
tripod and the HRT flux tube holders. The HRTs were provided with large area contacts
directly to the calorimeter and the mechanical support points of the tripod were
isolated from the HRTs to the extent allowed by structural rigidity considerations.
Two 30 $\Omega$ manganin wire heaters were wound in sections on three sapphire posts which were
attached to the circumference of the calorimeter 120$^{\circ}$ apart
to apply distributed heating. The heater sections were linked together with
superconducting wire. Two GRTs were also attached to the calorimeter. 

Near the end of the mission, before the shuttle landed, the helium cryogen was depleted,
causing the calorimeter to warm and become pressurized.  A removable burst disk allowed
the pressurized helium to vent into the vacuum can.  Because the calorimeter was
classified by NASA as a pressure vessel, it was required to be doubly fail-safe
against bursting.  This resulted in the addition of two fixed burst disks to the
calorimeter. These burst disk assemblies were designed to minimize the included
volume of helium and were fabricated from high conductivity copper to minimize
their thermal relaxation time. A pair of small heaters penetrated the calorimeter wall.
These were designed to cause vaporization of the superfluid when activated,
guaranteeing a thermodynamic path along the vapor pressure curve, in the unlikely
event that natural bubble nucleation in space was inhibited. 

%***************************
\begin{figure}
\resizebox{3.25in}{!}{\includegraphics{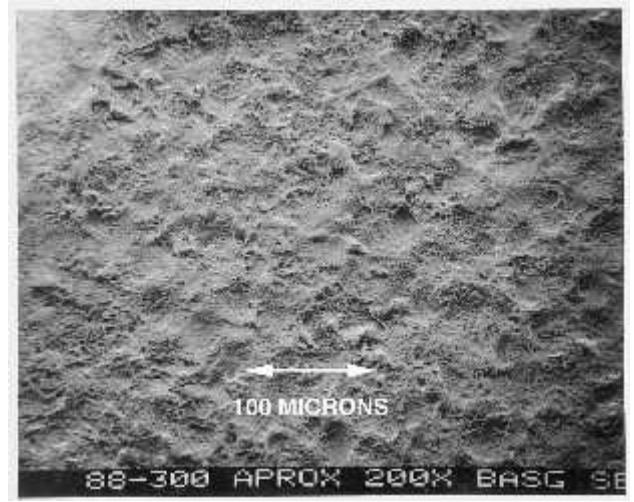}}
\caption{\label{fig:epsart2}Electron micrograph of the inside surface of the calorimeter
showing the scale of roughening prior to assembly.}
\end{figure}
%****************************************************************
%****************************************************************
\begin{figure}
\resizebox{3.25in}{!}{\includegraphics{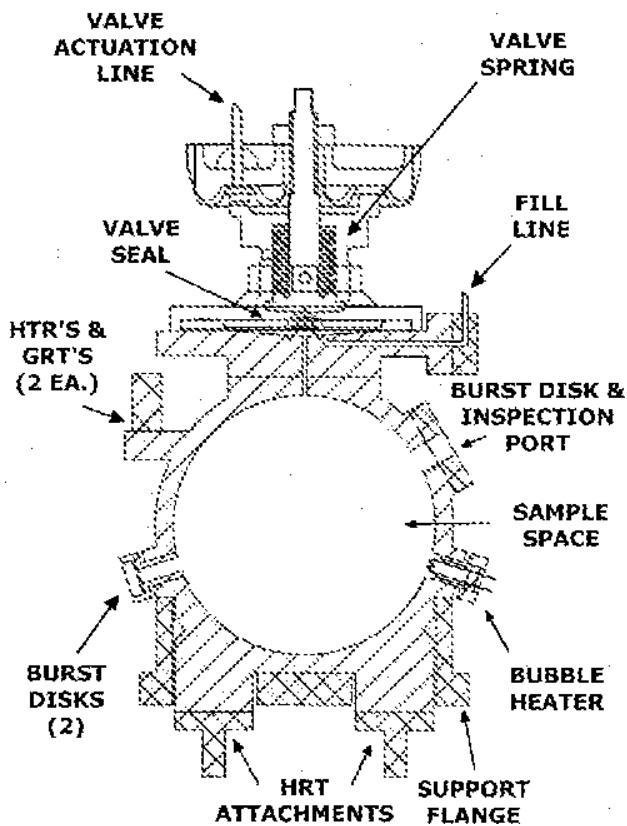}}
\caption{\label{fig:epsart3}Cross-sectional view of the calorimeter assembly, approximately
to scale.}
\end{figure}
%****************************************************************
After fabrication the inside of the cavity was lightly etched with an acid `bright dip' solution
to remove loose material, and inspected through the removable burst disk
hole using a fiber-optic light pipe  to check
for splatter from the welding. No problems were found. The cavity was then ultrasonically
cleaned and flushed with low residue ethanol
until the filtered waste no longer showed particle contamination. The valve assembly
and the final burst disk were then
installed in a class-10 clean room. A 2-micron filter was placed in
the fill line to guard against subsequent contamination.

\subsection{\label{sec:level2c}High Resolution Thermometers}
The thermometers developed for the experiment made use of the strong temperature dependence of the
magnetic susceptibility of copper ammonium bromide, CAB, and the high sensitivity of
SQUID magnetometers to resolve very small temperature changes. The basic design and performance
of the thermometers have been described extensively 
elsewhere, \cite{LipaPhysica1981Chui1992ChuiPRevLett1992} but for completeness we
include a brief description here, concentrating on those aspects relevant to the present experiment.
The external portion of the HRTs consisted of a niobium tube 0.83 cm internal diameter and 18 cm long,
with a constant magnetic field trapped along its axis. A 1.5 cm long cylinder of CAB was located near
the center of the tube. The magnetic moments of the salt molecules were partially aligned by
the magnetic field trapped in the flux tube. The magnetization of the salt changes with
temperature, inducing a d.c. current in a superconducting pick-up loop surrounding the cylinder,
which is measured by a SQUID magnetometer. A schematic view of the design is shown in Fig. 1
of ref. 28.
The field was generated by the solenoid wound on the vacuum can and was trapped in the
flux tubes early in the ground testing phase of the experiment. A reasonably comprehensive
idea of the behavior of the thermometers as a function of temperature and magnetic
field can be obtained by consideration of the Brillouin function applied to the
behavior of CAB, which has a magnetic transition near 1.83 K. \cite{Velu1976}

The pick-up coil and salt geometry were configured to maximize the sensitivity of the HRT.
The optimization sets the coil diameter so that the cross-sectional areas of the tube inside and
outside the coil are the same, when the salt fills the coil. The pick-up coil inductance
was matched to that of the SQUID input coil for maximum energy transfer, and the number of
turns around the salt was maximized for the highest 
sensitivity. \cite{SwansonLipa1994} The coil and salt were
placed in a sapphire holder which provided a high thermal conductivity link to the calorimeter
without introducing Johnson noise currents. One end of the holder was metalised and soldered
to a copper rod which thermally coupled the salt to the experiment. Copper wires embedded
in the salt were also soldered to this rod to provide additional thermal contact. The rod was
bolted to the base of the calorimeter using indium foil for thermal contact. The combined heat capacity of
the salt pill and thermal impedance of the copper rod give a time constant of about 1 second for heat
transfer between the salt and the calorimeter.  The flux tube was held by a chrome-copper sleeve which
was bolted to the support flange on the base of the calorimeter. The tube was a tight fit on the sapphire
holder to reduce microphonics in the pickup coil.

The niobium-titanium leads from the pick-up coil were tightly twisted together,
passed through a niobium-titanium
tube for shielding, and attached to a SQUID located elsewhere in the thermal
control system.  The tube was filled
with vacuum grease to promote thermal contact to the wires and to reduce microphonics.
A heater consisting of
a short piece of  0.005 cm diameter manganin wire was wrapped tightly around one lead near the SQUID input
terminals.  Applying a 10 mA current to this heater warmed the pickup loop wire above its superconducting
transition temperature in a fraction of a second, causing any persistent current
flowing in it to decay rapidly to zero. 
This capability was included because it was found that some available SQUIDs were
unable to operate with high currents in their input coils, limiting the dynamic
range of the HRTs to a temperature band $\sim0.4$ K wide. By activating this heater
with the calorimeter near the lambda point, a suitable operating range could be established.
During early testing two other effects were detected. It was found that the HRT calibration
was dependent on the temperature at which the current in the loop had been nulled.
We therefore chose a small set of specific temperatures at which to null the current,
as described in the calibration sections below. It was also found that a small
apparent thermal coupling existed between the calorimeter and the outer stages
of the thermal control system if a persistent current existed in the pickup loops.
This effect was probably due to slight changes in the superconducting penetration
depth of the pickup loop wire and shield with temperature at points where they
were thermally attached to the stages. To minimize this effect the currents were
again nulled within 1 mK of the transition.

To reduce the effect of the varying orientation of the HRTs with respect to the earth's
magnetic field, three layers of shielding were used.  The HRT flux tube itself had an
extremely high theoretical shielding factor at the location of the pickup loop, but
tests showed that this could be compromised by flux creep through the walls. \cite{MarekJap1987}
The practical shielding factor of this tube was expected to be $\sim5\times10^{9}$. 
A second superconducting shield was added around the lower portion of the instrument
vacuum can with a theoretical shielding factor of about 200.  The third shield was a
layer of moly-permalloy, \cite{WilliamsCorp} a high permeability material, placed around the outside
of the cryostat to gain a further factor of 100.  Tests of the complete assembly showed no
detectable signal to a level of $\pm 3\times10^{-4}\Phi_{0}$ for magnetic fields
of $10^{-4}$ T applied externally, where $\Phi _{0}$ is the quantum of magnetic flux.
In traversing an orbit, the experiment was subjected to a variation of about $\pm 5\times10^{-5}$ T
from the earth's magnetic field. 

All the HRTs were grouped together below the calorimeter for convenient charging
by the solenoid on the vacuum can. With the winding energized, the HRTs were
warmed above 10 K and then cooled back to 4.2 K where the field was slowly
reduced to zero.  During this period the SQUID magnetometers remained near
their operating temperature allowing the field trapping process to be monitored.
It was found possible to detect the superconducting transition in the flux tubes
and in the pickup coil wire. The field-to-current ratio of the solenoid
was approximately $1.1\times10^{-2}$ T/A 
and the field variation over the volume of
the HRTs was measured to be $< 1\%$. The trapped field level was selected
as a compromise between the effects of digitizer nonlinearities and HRT
drift performance. For a field of 10$^{-2}$ T trapped in the flux tubes,
the sensitivities of the HRTs were typically in the range 2 - $3\times10^{6} \Phi_{0}$/K.
Their noise performance is discussed below.

\subsection{\label{sec:level2d}SQUID Magnetometers}
Four r.f. SQUID magnetometers were attached to an independent single-stage thermal platform
supported from the lid of the vacuum can. This arrangement was necessary to
minimize the effect of the temperature coefficient of the
SQUIDs \cite{LT17LuuLipa1984} on the HRT output, and to
maintain the critical currents of the sensors within their operating range.  The SQUID platform
was stabilized to $< 10^{-4}$ K using a servo similar to those on the outer stages of the thermal
control system.  The SQUIDS were housed in individual cylindrical
niobium shields \cite{RigbyChui1990} closed
at one end, with magnetic attenuation factors of $\sim5\times10^{9}$. The open end of the tube was filled
with a threaded niobium plug containing a small opening for the leads.  Each assembly was placed
inside a close-fitting 
Cryoperm \cite{Germany} shield equipped with a degaussing coil.  Before operation
the SQUIDs and shields were warmed from their operating temperature of 4.5 K to above 10 K to
reduce any internally trapped fields.  They were then cooled slowly through the superconducting
transition to minimize fields generated by thermoelectric currents.

The outputs of the SQUIDs were connected via triaxial cables to r.f. pre-amps and associated circuitry
attached to the exterior of the cryostat. No modifications were made to the commercial 19 MHz r.f. SQUID
sensors, \cite{BiomagTech} but the electronics were rebuilt to improve operation in space. 
They were strengthened
to survive launch, repackaged to improve their thermal performance in vacuum and modified for
computer-controlled tuning.  The temperature of the pre-amps was controlled to $\pm 0.1^{\circ}$C to
minimize d.c. offsets that would mimic temperature changes of the HRTs, and the A/D converters
reading the outputs were controlled to about $\pm 1^{\circ}$C. In operation, the SQUID magnetometers
give voltage outputs in terms of the magnetic flux change at the input, from which the corresponding
changes in temperature of the salt pills can be inferred.
The noise floor of the SQUIDs was on the order of $10^{-4}\Phi_{0}/\sqrt{\text{Hz}}$.
The output bandwidths were limited to 0.16 Hz by double pole active filters.

\subsection{\label{sec:level2e}Adsorption Pump}

During early ground testing it was found that the launch vibration was capable of warming
the HRTs on stage 4 to $>$ 15 K and the calorimeter to $>$ 5 K.  Either of these temperature excursions
could have ruined the experiment as there was no provision to re-magnetize the HRT
flux tubes in flight and
the pressure of the helium in the calorimeter could have exceeded the set point of the
removable burst disk.
To reduce the temperature rise we placed approximately 0.3 mole of $^{3}$He exchange gas in the vacuum can
prior to launch.  This gas greatly increased the thermal coupling between the inner
stages of the instrument
and the vacuum can. 
Additional shake tests simulating the launch vibration indicated that the gas
limited the temperature rise to $< 0.1$ K.  After launch, most of the gas was evacuated to
space through the instrument vacuum pumping line. To obtain a high vacuum in the
instrument during the heat
capacity measurements an adsorption pump containing $\sim 50\text{ cc}$ of activated charcoal was
located in the lid of the vacuum can.
The charcoal container was thermally isolated from the lid allowing it to be
heated to $\sim 40$ K without significantly warming other components.
The entrance to the pump was baffled with a gold
coated copper disk which was thermally attached to the lid. This was needed to reduce
the thermal coupling between the pump and stage 4 of the thermal control system.

To estimate the pressure in the system we monitored the power P$_{4}$ dissipated in stage 4
of the thermal control system when the temperature difference T$_{4}$ - T$_{3}$ was
held at a specific value. During ground testing it was found that for pressures
$> 10^{-9}$ torr the additional power needed
to compensate for the gas conduction was detectable. We roughly calibrated the rise
in P$_{4}$ against pressure using an external leak detector, allowing for thermomolecular effects.
With no exchange gas in the
system the pressure was typically $<10^{-9}$ torr, the limit of the measuring technique. 
When exchange gas was introduced, the pump-down time constant was on the order of a day,
dependent on the vacuum can wall temperature. After about two days, the pressure
in the vacuum can was generally low
enough to perform high resolution measurements.  However, the residual gas had small but
detectable effects on the HRTs which were modeled and corrected for as described in the
analysis section below.

\subsection{\label{sec:level2f}Cryostat and Electronics}
The cryostat system used to maintain low temperatures was 
inherited from a previous flight program. \cite{MasonAdvCryoEng1980}
No significant modifications were made to the cryogenic portions. However, the external supports
were reconfigured to operate with a magnetic shield which was added to the exterior.  The cryostat
and its performance have been described elsewhere. \cite{PetracLuchik1994} 
The liquid helium in the cryostat was
cooled to $\sim 1.7$ K during the flight by venting it to space through a sintered metal plug
designed to contain the superfluid.

The instrument electronics consisted of nine 4-terminal a.c. resistance bridges,
four r.f. SQUID controllers, four precision heater drivers with D/A converters and
five analog servo circuits for temperature control. A number of digital switching
circuits were also included.  Five additional temperature controllers were used within
the electronics to decrease the sensitivity of critical components to variations in the
ambient temperature. The electronics were operated by a small PC-style computer with a
286 microprocessor and a basic software duty cycle of 1 Hz synchronized with a shuttle timing
circuit.  Once per cycle a 400 byte telemetry package was 
passed to a communications computer which also processed any
commands sent from the ground.  A third computer interfaced
with the facility hardware to provide cryostat housekeeping functions. Most of the instrument
tasks were broken into segments, limiting the amount of code executed in each cycle to
avoid interference with the communication task. 

The most critical circuits for heat capacity measurements were the heater current monitors
and the A/D converters on the outputs of the SQUIDs.
These converters had nominal 16 bit resolution, but were found to have
occasional non-monotonic behavior of the output vs. input.
Separate measurements indicated that the converters were likely to have $< 2$ bits
deviation from a straight line fit. Since the HRT noise level was typically 30 bits pp,
substantial averaging over the bit errors occurred, reducing the corresponding
temperature measurement errors significantly. For the heater current monitors
the deviations from a linear fit were bounded by a $\pm 0.01\%$ power measurement
band over the full range of powers used in the experiment after corrections for the
gains and offsets of the A/D converters. 

The driver circuits for the heaters on the calorimeter had dedicated timers to allow precise control
of the energy within a given heat pulse. The pulse time was
accurate and stable to $< 2$ parts in $10^{6}$.
The rise and fall times were $<$ 60 $\mu$s duration, giving a maximum correction 
of $0.0024\%$ for the minimum 5 sec pulse length which was neglected.
The heater driver circuit had 6 power ranges to accommodate a wide range of
temperature step sizes while operating the A/D converters at a high bit setting.
To allow automatic data collection a software routine was written to select the
pulse length, power range and D/A setting optimum for a given temperature step.
Manual operation was also possible.

Two plastic scintillators with photomultiplier tubes were attached to the exterior of the electronics.
These were used as charged particle monitors (CPMs) for correlation with effects in
the instrument. The voltage applied to each photomultiplier tube was programmable
over a wide range to control the gain. The output of each CPM was threshold-detected at
two levels, approximately 10$\%$ and 90$\%$ of saturation.  The number of counts
accumulated in each second for each channel of both detectors
was included in the telemetry stream.

Two 3-axis accelerometers \cite{RogersAdvSpace1998} with
sensitivities of $10^{-6}$ g were attached to the outside shell of the cryostat. The signals were
sampled at 250 Hz and digitally filtered to a 100 Hz bandwidth.  The vector amplitude of
the acceleration was calculated and the peak value in each 1-second interval was output
through a separate telemetry system to aid in characterizing the vibration environment.
Occasional segments of raw data were also downlinked.

\section{\label{sec:level3}PRE-FLIGHT OBSERVATIONS \protect\\}
Prior to its installation in the flight cryostat, the instrument was operated in the laboratory. 
In this period the volume of the calorimeter was determined by measuring the mass of gas extracted
after a low temperature fill.
In conjunction with data on the size of the small gas bubble left in
the calorimeter near the lambda point,
this information allowed us to determine the mass of the flight sample. The heat capacity of the empty
calorimeter was measured, and the heater circuits were calibrated with board level and dummy load tests.
Upper limits were set on variations of the stray power levels in the calorimeter heaters, and the GRT
calibrations were checked in situ against a rhodium-iron 
thermometer \cite{LakeshoreModelSN} (RIT) calibrated by NIST.

After the instrument was installed in the flight cryostat, the flight sample was sealed in
the calorimeter and the system was maintained at or below 4.2 K for an 18-month period while
it underwent environmental testing and integration with the shuttle. During this period
a number of checks were made to verify its correct operation.  These included additional heater
stray power measurements, a magnetic field susceptibility test, an electromagnetic susceptibility
test, HRT calibrations, and lambda point observations on the GRT temperature scale. 
The instrument was shake-tested and the electronics were operated in a thermally controlled vacuum
chamber. Heat capacity measurements were also made below the lambda point.
The following subsections summarize the most critical 
measurements performed during the pre-flight period.

\subsection{\label{sec:level3a}Thermometers}
The GRTs on the calorimeter were calibrated 
by the manufacturer \cite{LakeshoreKrange} and
in the laboratory against the vapor pressure of $^{3}$He and the RIT
using the ITS-90 \cite{PrestonMetro1990} temperature scale.
For these measurements a
specially built apparatus was used. The GRT resistance calibrations
were performed using an a.c. bridge technique \cite{RubinSciInst1972}
with a seven-decade
resistance standard \cite{ElectroSciIndModelSN} as the reference. The $^{3}$He calibration
was performed using high purity gas and included corrections for
thermomolecular and hydrostatic effects. In the range 1.8 to 2.4 K
temperatures derived from the $^{3}$He calibrations deviated from
those obtained from the GRT manufacturer's calibrations by about
0.5 mK on average, with a peak deviation of 0.7 mK. The RIT measurements
were consistent with the $^{3}$He calibrations to within 0.3 mK over
the interval 1.72 to 2.63 K. However it was found that the lambda
transition temperature of a small sample of $^{4}$He was within
0.1 mK of that derived from the $^{3}$He calibration, leading us
to prefer this scale for the experiment. The resistance values of
the GRTs were converted to temperature using the formula: \cite{LeungCryog1979}
%=============================================================
%\begin{equation}
\begin{eqnarray}
\text{log T} = \text{a}_{0} + \text{a}_{1}\text{log R} +
\text{a}_{2}(\text{log R})^{2} + \text{a}_{3}(\text{log R})^{3}   
\end{eqnarray}
%\end{equation}
%============================================================ 
where the coefficients a$_{i}$ were determined from least squares fits
to the $^{3}$He calibration data.  Between 2.0 and 2.4 K the deviations
from the fits showed no significant trend exceeding their 15 $\mu\text{K}$ rms scatter.
The bridges in the flight electronics were calibrated by the manufacturer
(Ball Aerospace) and checked against our GRT bridge resistance standard
mentioned above, which was separately checked by the JPL standards group.
Corrections of up to $0.4$ $\Omega$ from the nominal resistance values were
stored in lookup tables in the flight computer.  After the GRTs were installed
in the flight instrument they were recalibrated against the RIT, which was
then removed.  It was found that the resistance vs. temperature scale factors
of both GRTs appeared to have shifted by about 0.06$\%$ relative to that
of the RIT, but by less than 0.01$\%$ relative to each other. This discrepancy
appeared to be associated with a thermal anchoring problem with the RIT.
Since measurements of the apparent temperature of the lambda transition
also discriminated against the RIT, it was assumed that the GRT calibrations
had not changed during installation. In either case, the absolute uncertainty
in the values of (dR/dT)$_{\lambda}$ appeared to be limited almost
entirely by the  0.1$\%$ uncertainty in ITS-90.

Over the course of the experiment it was found that the apparent lambda
transition temperature of one of the GRTs ($\#$2) drifted slightly.
While this drift was less than 0.2 mK over an 18-month period, it
was substantially greater than the 30 $\mu$K drift seen in GRT $\#$1.
Later measurements on a second flight experiment to measure the specific 
heat of confined $^4$He \cite{LipaPhyRevLett2000} confirmed the
continued slight drift of GRT $\#$2. We also found that the noise
level of bridge $\#$2 was a factor of 2.3 higher than that of $\#$1.
If we treat the drift as a scale factor change in the R(T) curve,
we can estimate the change to be $\sim$ 0.2 mK/T$_{\lambda}$, or
0.01$\%$ which is negligible compared to the overall uncertainties of the experiment.

The HRTs attached to the calorimeter were calibrated against the GRTs over
a 100 mK range just below the lambda point. This was done by repetitively
bringing the apparatus into thermal equilibrium, recording the resistance
and flux values, and then stepping the calorimeter temperature in
increments of $<$ 1 mK. The HRT flux readings vs. GRT temperatures
were fit over the full range using a Curie-Weiss model of the HRT salt behavior.
The function used was
%=============================================================
%\begin{equation}
\begin{eqnarray}
\phi - \phi_{\infty} = \text{H}_{0}(1 + a\phi)(T/T_{c} - 1)^{-\gamma}                                    
\end{eqnarray}
%\end{equation}
%============================================================ 
where H$_{0}$, $\phi_{\infty}$ and $a$ were constants determined
in the fit, and the flux $\phi$ was measured from the temperature at
which the persistent current in the pickup loop was nulled.
The term proportional to $\phi$ on the right-hand side allows for the variation
of the magnetic field with current in the pickup loop as the temperature
is changed. We chose T$_{c}$ = 1.8302 K and $\gamma = 1.2736$ based
on the data of Velu et al. \cite{Velu1976} for CAB. Typical values obtained for
the other parameters in Eq.(5) were:
$\text{H}_{0} = 1.2158\times10^{5}$ $\Phi_{0}$,
$\phi_{\infty} = -1.5686\times10^{6}$ $\Phi_{0}$,
and $a = -2.227\times10^{-7}/\Phi_{0}$.
In our earlier analysis \cite{LipaPhyRevLett1996} we used
a third order polynomial in place
of Eq.(5), fit over a 50 mK range. However we recently
found that in the range $3\times10^{-3} < t < 10^{-2}$ the heat capacity
results were somewhat sensitive to the values of the 
second and third order coefficients.
This effect led to a dependence of $\alpha$ on the details of the HRT
calibration procedure that exceeded our goal of no more than
$10^{-4}$ for systematic bias. The function in Eq.(5) gave
results that were more stable in this outer region of $t$, and gave a better
fit to the calibration data over a wider range of $t$. This can be inferred from
Fig. 4 which contrasts the two calibration procedures.
It shows the GRT data and
various calibration curves as deviations from the best fit to
Eq.(5) over 100 and 50 mK ranges. If we attempt to fit the data over the 100 mK
range with a 3rd-order polynomial, we obtain the dot-dash curve: clearly a
higher order polynomial is required for an adequate fit over this range.
If we restrict the 3rd-order polynomial fit to 50 mK 
from $\text{T}_{\lambda}$ as was done
in the previous analysis of the flight data, we obtain the dashed curve.
While this fit represents the data well in this restricted range, it
rapidly deviates from the data set outside the fit range and more
importantly it is overly sensitive to small local trends in the data set.
In particular, the critical exponent, alpha, is sensitive to the curvature
seen in the deviation plot of the 3rd-order polynomial in the range 0 to
20 mK below $\text{T}_{\lambda}$. 

%****************************************************************
\begin{figure}
\includegraphics{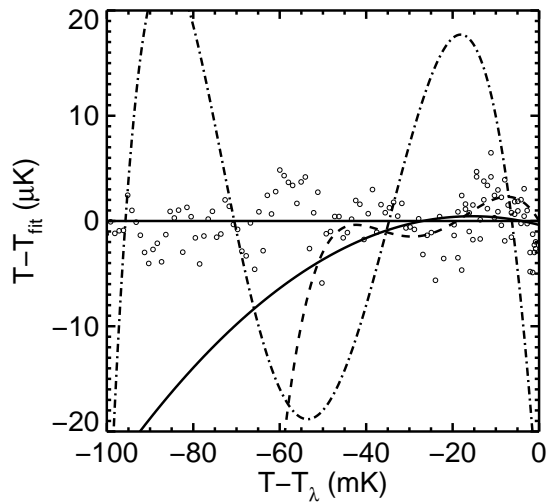}
\caption{\label{fig:epsart4}Deviations of the GRT $\#$1
data and various calibrations from the best
fit of the HRT data to Eq.(5). The open circles are the GRT data, the
dot-dash curve is a 3rd-order polynomial fit to the entire data set, the
dashed curve is a 3rd-order polynomial fit in the range 0 to -50 mK, and
the solid curve is a fit to Eq.(5) over the same restricted range.}
\end{figure}
%****************************************************************

In the flight experiment, calibration
data were only available over a 50 mK range so it is important to
demonstrate that a Curie-Weiss model fitted over this restricted range
will still represent the data. The deviation of the Curie-Weiss model
fitted over the 50 mK range from the fit over the full range is shown by
the solid curve. It is clear that over the critical 20 mK range where heat
capacity measurements were performed, the restricted Curie-Weiss model is
superior to the 3rd-order polynomial fit. The curvature in the deviation
plot over this range is substantially decreased, resulting in a reduction
of systematic errors in the determination of alpha. This new calibration
procedure is chiefly responsible for the difference between the value of
alpha reported in ref. 11 and that reported here.

The statistical uncertainties of (d$\phi$/dT)$_{\lambda}$ 
for the HRTs were $<$ 0.004$\%$ and
no significant deviations were apparent between 2.12 and 2.177 K. It was found
that the deviations from the fit tended to be large when a high bit was
switched in the 16-bit reference resistors of the GRT bridges.
We also found that the relays used for the switching often had
a higher-than-specified resistance immediately after a contact had been closed.
The calibration of the HRTs included the weighted data from both
GRTs as well as 11 extra parameters for each GRT to model any additional
contact resistance introduced by switching. The fit also included the
constraint that the contact resistances are the same for the calibration
of the two HRTs to ensure that there is only one GRT scale.
These resistances were
found to be $< 0.33$ $\Omega$ in all cases, however, their inclusion
reduced the scatter in the GRT readings by about a factor of two. The curve
fitting process left a slight residual relative error in the temperature
scales of the two HRTs. Since this resulted in slightly different heat
capacity values depending on which HRT was used, we performed a second
fit of the HRT data to a temperature scale defined by the average of
the two.  

To obtain reproducible calibration coefficients for the HRTs it was necessary
to control the value of the persistent current flowing in the pickup loops.
This current can perturb the field applied to the salt pill as shown in Eq.(5),
changing its magnetization. We minimized this effect by setting the current to
zero at fixed reference temperatures, using the heater attached to the HRT leads.
The ground heat capacity data set was collected with the heaters activated while
the HRTs were at 2.076 K. 

The low frequency noise level of the HRTs was about $2.5\times10^{-10}\text{ K}/\sqrt{\text{Hz}}$,
which is close to the value predicted by the fluctuation-dissipation theorem
of Callen and Welton \cite{CallenPhysRev1951} applied to temperature fluctuations of a
paramagnetic salt. The noise measurements have
 been described elsewhere. \cite{QinChuiCryogen1996,LipaPhysica1981Chui1992ChuiPRevLett1992}
Wide band random vibration tests were conducted to determine the various
instrument sensitivities. It was found that the effect on the HRT noise
was very nonlinear, being substantial above a r.m.s. vibration level of
$10^{-2}$ g but relatively small below $7 \times10^{-3}$ g.
With the expected acceleration levels below $2 \times10^{-3}$ g except
for occasional transients, little effect on the thermometry was anticipated.

\subsection{\label{sec:level3b}Heater Power}
There are two aspects of the calibration 
of the calorimeter heater power, P$_{5}$, that are relevant for the heat capacity results:
measurement of absolute power parameters and an estimate of stray heater power. The heater
resistance was measured at low currents to within 0.02$\%$ and the heater current monitor
circuit was calibrated to 0.01$\%$ as mentioned earlier.  The temperature coefficient of
the heater resistance was also measured and applied in the analysis. The heater parameters
could not be checked accurately in flight, but the heater current monitor circuit was tested
in a thermal/vacuum chamber which simulated the space environment. Its calibration was found
to be stable to $\pm 0.01\%$ over the temperature
range from $-10^{\circ}$ to $+30^{\circ}\text{ C}$. During
the experiment, the electronics box temperature
was in the range $18^{\circ}$ to $23^{\circ}\text{ C}$. 

Residual or stray power dissipated in the heater can affect the heat capacity results
if it varies between the heater 'on' and 'off' states. For example this could be
caused by bias currents in the active elements of the circuit
or by high frequency EMI that changes its level when the heater is switched. To reduce
this possibility the heaters were terminated with resistive loads similar to those
presented by the current supply circuits when switched to the 'off' state. Pulses
were applied by first switching the heater power range and the associated D/A converter
to the desired settings, then switching the heater from the resistive load to the
circuit using CMOS switches controlled by a precision timer.  By performing the switching 
from the 'off' to the 'on' state and leaving the D/A converter output at zero volts
while monitoring the heating rate of the calorimeter, we were able to set
a limit on the stray power change of $< 10^{-11}$ W. 

A more comprehensive, though less precise test of the entire heat input circuit can be
performed by comparing heat capacity values as a function of power. A special sequence
of heat capacity measurements which consisted of a set of pulses covering the
range of available power was performed at a temperature where the heat capacity is only
weakly temperature dependent, $\sim$ 20 mK below the transition.  Since the sample heats
slightly as the pulse sequence progresses, a symmetrically arranged reverse sequence
was also performed. By fitting the heat capacity results at the same power with a model
function, the effect of the temperature dependence of the specific heat could be minimized.
These measurements were made on a number of occasions and typical results are shown
in Fig. 5.  From the measurements we concluded that the apparent change in the stray
heater power during pulses was $-9.4 \times10^{-11}$ W.  Since this was significantly higher
than the value obtained in the zero input voltage switching test above, we concluded that
most of the effect was due to a zero offset in the measurement circuit. It was also found
that the apparent heat capacity decreased by $\sim 0.17\%$ at the highest power used,
possibly due to self-heating in the heater assembly. These effects were modeled with a
four-parameter function and corrections were applied to all pulse power estimates.
For P$_{5}$ $> 10^{-5}$ W the uncertainties in the fit were $< 0.005\%$.

%****************************************************************
\begin{figure}
\includegraphics{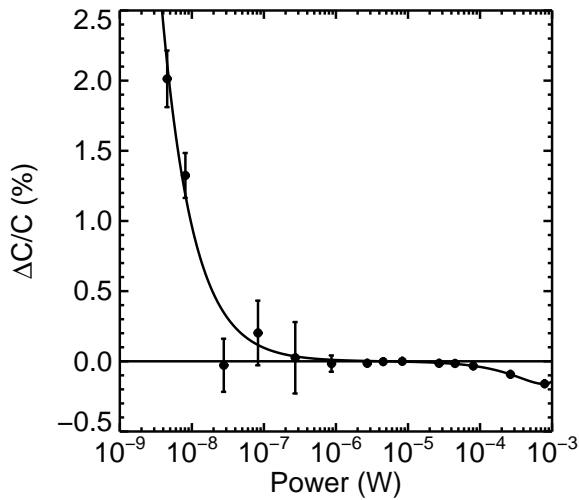}
\caption{\label{fig:epsart5}Dependence of the apparent heat capacity on the power
dissipated in the heater as measured during pre-flight testing. Curve shows model fit.}
\end{figure}
%****************************************************************

Mechanical vibration was also expected to be a variable heat source
during the flight experiment. From the vibration testing mentioned
earlier we derived a calorimeter heating sensitivity of $\sim$ 0.2 W/g$^{2}$/Hz using
wide band random vibration input with power spectral densities
in the range $3\times10^{-6} - 2\times10^{-7}$ g$^{2}$/Hz. With a swept
sine wave input we determined that about 95$\%$ of the heating
occurred in a narrow frequency band near 55 Hz. Since spectral
densities of 10$^{-9}$ - 10$^{-10}$ g$^{2}$/Hz were expected
in the mission, some low level heating from this source was anticipated.
Unfortunately it was not possible to add vibration isolation
to the apparatus due to cost.

\subsection{\label{sec:level3c}Sample Mass}
The apparatus used to fill the calorimeter did not allow an accurate measurement
of the quantity of helium sealed inside. Instead, we measured the sample mass by
a two-step process which first involved measuring the calorimeter volume and
the mass of a reference sample.  The mass of the reference sample was measured
during its extraction from the calorimeter. The volume was estimated by taking
advantage of the very well known \cite{KerrAnnPhys1964} temperature dependence of the density of
helium along the vapor pressure curve near the lambda point.  If the calorimeter
is between 99.5 and 100$\%$ full at the lambda point, 
the gas space can be reduced to zero by cooling to a 
temperature $2.18 > \text{T}_{B} > 1.8$ K.  Above T$_{B}$ the
thermodynamic path is close to that of the saturated vapor pressure, while below
T$_{B}$ it is at constant volume.  The passage through T$_{B}$ can be detected by
the small but sharp change in the specific heat due to the change in the thermodynamic path.
We were able use T$_{B}$ as a very sensitive indicator of the calorimeter fill
fraction at the lambda point.  The number of moles in the reference sample
and the molar volume at T$_{B}$ can easily be used to derive the calorimeter volume,
resulting in a value of 22.332 $\pm$ 0.01 cc.
Combining this value with that of T$_{B}$ for any other sample is then sufficient
to determine its mass. An error of 1 mK in T$_{B}$ corresponds to a mole number
change of $< 0.0015\%$ which is negligible compared with the uncertainty in the calorimeter volume.

The variation of the calorimeter volume from fill to fill was extremely
small due to the design of the fill line valve. 
Closure was accomplished with a metal-to-metal
seal with constant applied force.
On the calorimeter side
of the valve the variable volume was $< 2 \times10^{-3}\text{ cc}$ during
the seal indentation process.
It is unlikely that the variation in volume 
after achieving a seal exceeded $10^{-3}\text{ cc}$, which is
negligible compared to the total volume. After the calorimeter was
filled with the flight sample, T$_{B}$ was again determined. The corresponding helium mole
number was $0.81314 \pm 0.0004$ and the gas space at the lambda point was estimated
to be $0.0558 \pm 0.0007$ cc.

\subsection{\label{sec:level3d}Empty Calorimeter Heat Capacity}
The heat capacity of the empty calorimeter was determined by heat pulse measurements
and indirectly from its relaxation time plus the thermal resistance to stage 4. 
Both methods gave consistent results. The heat capacity of the empty calorimeter
at the lambda point was found to be $0.0441 \pm 0.0005$ J/K, which agrees with a
rough estimate based on the knowledge of its components.  The temperature dependence of
the heat capacity was measured to be 0.019 $\pm 0.003$ J/K$^{2}$, which is negligible
over the 20 mK range of interest.  Nevertheless, a correction for this effect was applied.

\subsection{\label{sec:level3e}Ground Specific Heat Measurements}
The basic method of measuring heat capacity was to establish a suitably low drift
rate for the calorimeter temperature, apply a heat pulse of 5 - 35 s duration and then
wait for the drift rate to return to a low value.  The heat capacity was then computed
in the usual way from the rise in T$_{5}$ and the energy dissipated. Since the time
constant for the relaxation of T$_{5}$ to T$_{4}$  was $\sim 10^{6}$ s, simple
linear extrapolation to the center of the pulse was generally sufficient to obtain
the temperature rise. A least squares straight line fit was made to 50 - 120 s of
drift data on each side of the pulse and the temperature change and value at the
center of the pulse were estimated. The fitted data did not include the first 6 to
30 seconds of data after the heater was turned off.  This was done to allow for
the settling of small transients due to the warming of the calorimeter support
structure and the Kapitza boundary resistance, which slightly distorted the
measurements. Near the lambda point it was necessary to reduce the size of the
temperature steps to maintain accurate measurements of the local heat capacity
and to achieve the desired resolution of the singularity. This resulted in a
deteriorating signal/noise ratio as the transition was approached. 

A software routine was used to generate a sequence 
of measurements that adjusted the size of the steps
in a predetermined fashion and established the flux reading of the HRTs that
corresponded to T$_{\lambda}$. The routine started by adjusting T$_{4}$ to reduce
$\dot{\text{T}}_{5}$
below 10$^{-10}$ K/s. A heat capacity measurement with a moderate
energy input was then performed and the specific heat was calculated. The remaining
temperature increment to the lambda point was then estimated by comparing the
measured specific heat with a functional form based on previous experiments. 
To avoid ambiguity this routine was used only below the transition. For subsequent
measurements in the sequence the routine used the estimate to determine the energy input
to the heater to give a temperature step $\Delta$T equal to a specified fraction of the
remaining interval to the transition. Prior to each pulse
$\dot{\text{T}}_{5}$ was measured and
adjusted to $ < 10^{-12}$ K/s or $2\times10^{-4}\Delta$T K/s, whichever was greater.
After each pulse the estimate was updated. The routine allowed rapid collection of
data approximately equally spaced on a logarithmic temperature scale, a distribution
desirable for the final curve fitting task. It also maximized the time available for
repeat measurements at high resolution that were necessary to improve the signal/noise ratio.  

When used far below T$_{\lambda}$ the routine also generated a string of alternating
low and high power pulses. The low power pulses were performed to provide results
over a wide range of $t$ at a fixed low power in case unexpected power-dependent
effects were detected in the analysis. The typical step size for these pulses
was 1 $\mu$K and the heater power was $8\times10^{-6}$ W. The high power pulses
started with the maximum power available, $8\times10^{-4}$ W, and were designed to warm
the calorimeter about $10\%$ of the remaining distance to the transition.
For $t \sim 0.01$, a pulse duration of 35 s was used, decreasing as the
transition was approached. When the desired temperature step fell below 300 $\mu$K,
the high power pulses were discontinued.

The pulse method gives the specific heat integrated over $\Delta$T, which differs
from the value at the midpoint temperature
due to the curvature of the specific heat
function. A correction
given by $\Delta\text{C} = \Delta\text{T}^{2}(\partial^{2}\text{C}/\partial\text{T}^{2})/24$  can be made
if the function is known. To evaluate the second derivative we used
the specific heat function given in ref. 5.
This correction was $<0.07\%$ of the specific heat for all pulses. 

Given the total specific heat of the sample, C$_{T}$, some additional small
corrections are needed before a fit with a theoretical expression is performed.
The presence of a gas space introduces small corrections to the specific heat
primarily via the expansion coefficient and the latent
heat. \cite{KellersDuke1960,AhlersPRevA1971} The most
important effect is due to the changing volume of the gas space with temperature
and the correction is given by 				   
%=============================================================
\begin{equation}
\text{C}_{S} = \text{C}_{T}/\text{n}_{l} + \alpha_{s}\text{L}\rho_{g}/\rho_{l}
\end{equation}
%============================================================ 
where $\text{C}_{S}$, $\alpha_{s}$, n$_{l}$, $\rho_{l}$ and L are
the specific heat, expansion coefficient,
mole number, density and latent heat of the liquid, and $\rho _{g}$ the density of the gas.
This correction
term is independent of the size of the gas space,
but depends on its rate of change with temperature.
A much smaller term, dependent on the volume of the gas, was also carried
in the analysis. Another correction term is from the conversion of the specific
heat along the vapor pressure curve to the specific 
heat at constant pressure. \cite{Ahlersbook1976}
The primary correction term is given by				  
%=============================================================
\begin{equation}
\text{C}_{p} = \text{C}_{S} + \text{T}v\alpha_{p}(\partial\text{P}/\partial\text{T})_{svp}                                            
\end{equation}
%============================================================ 
where $v$ is the molar volume of the liquid, $\alpha_{p}$ is the isobaric
expansion coefficient \cite{MuellerPhyRevB1976} and 
$(\partial\text{P}/\partial\text{T})_{svp}$ is the slope of the vapor
pressure curve.
Other corrections for conversion to a constant pressure path were found to be negligible.
The corrections to the specific heat in Eqs. 6 and 7 are small, $< 0.1\%$, but
not entirely negligible in the experiment.
The change of $\alpha$ due to these corrections was found to be $<10^{-5}$.

With the heat pulse method the fractional uncertainty in C is just
$(\sigma_{\text{C}}/\text{C})^{2} = 
(\sigma_{\text{Q}}/\Delta\text{Q})^{2}+(\sigma_{\text{T}}/\Delta\text{T})^{2}$
where the $\sigma$'s are the corresponding uncertainties
in the energy input and temperature step measurements.
As described above, P$_{5}$ was modeled to 
$\sigma_{\text{Q}}/\Delta\text{Q} \sim 0.005\%$
at the higher power ranges while the overall
accuracy was
$\sigma_{\text{Q}}/\Delta\text{Q} \sim 0.01\%$. 
This was typically less than the fractional
uncertainty in the temperature step, $\sigma_{\text{T}}/\Delta\text{T}$,
especially near the lambda point. To obtain good precision at high resolution it was therefore
necessary to average as
many measurements as possible. The absolute uncertainty of the results depends
not only on the noise but also on
the accuracy of the heater and thermometer calibrations. 
However, the uncertainty of the best fit
exponent is not dependent
on the overall scale factors that affect the specific heat: instead
it is set by the noise level of the data, its range,
any temperature dependence of the scale factors, and the fidelity of the curve fitting function. 
The statistical uncertainties we used for the results were based on the
formula above which gave values as low
as $\sigma_{\text{C}}/\text{C} = 0.008\%$.  We note that this does not include the $\sim 0.1\%$ systematic
uncertainty from ITS-90, nor the $0.05\%$
uncertainty in our determination of the sample mole number.
We estimate the absolute uncertainty in our specific
heat values to be about $\pm 0.2\%$.

%****************************************************************
\begin{figure}
\includegraphics{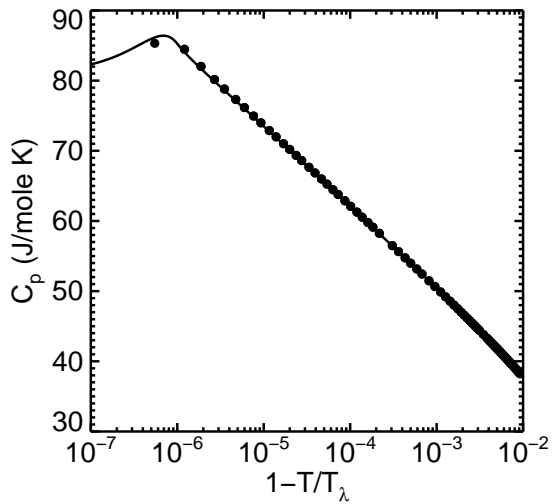}
\caption{\label{fig:epsart6}Ground specific heat results  below T$_{\lambda}$ 
on a log-linear scale. Line:
gravity rounded model.}
\end{figure}
%****************************************************************
%****************************************************************
\begin{figure}
\includegraphics{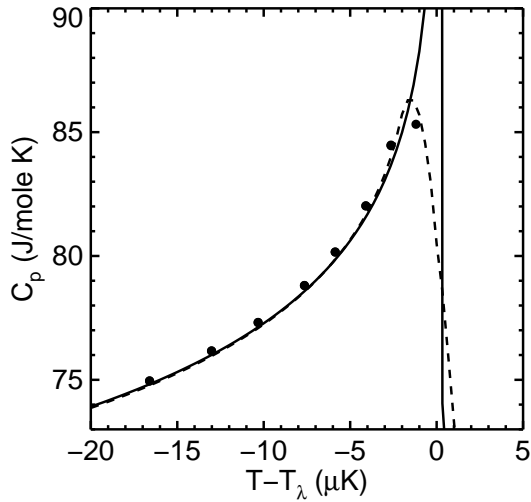}
\caption{\label{fig:epsart7}Ground specific heat results on a linear scale close to the
transition. Solid line: zero gravity function; broken line: gravity rounded model.}
\end{figure}
%****************************************************************

The results obtained for the specific heat prior to the flight are shown in 
Fig. 6 on a semi-logarithmic
scale. Close to the transition the curve is distorted by the effect of gravity 
on the sample. This leads
to a rounding of the singularity over a range of about 5 $\mu$K.
A magnified view of this region is shown
in Fig. 7 on a linear temperature scale. The solid lines indicate the 
expected behavior in the absence of
gravity based on our current fit to the flight data.
The broken line shows the expected curve in the
presence of gravity. 
The transition temperature was estimated by analyzing the thermal overshoot that
occurs after a pulse which causes the helium to enter
the two-phase region where normal and superfluid coexist within the calorimeter. 
We obtained an uncertainty of $\pm54$ nK in the temperature at which normal 
fluid was first stable within
the calorimeter. In Fig. 8 we show the data as deviations from a zero-gravity 
reference function given below by Eq.(9), with parameters
determined by a fit to the flight data. It can be seen that there is an offset 
of about $0.07\%$ between
the data and the function. We believe this is due to slight changes in the calibration of the HRTs and
heater circuit between ground and flight. Otherwise there is good agreement with the general behavior
of the function up to the region where gravity effects become significant.
%****************************************************************
\begin{figure}
\includegraphics{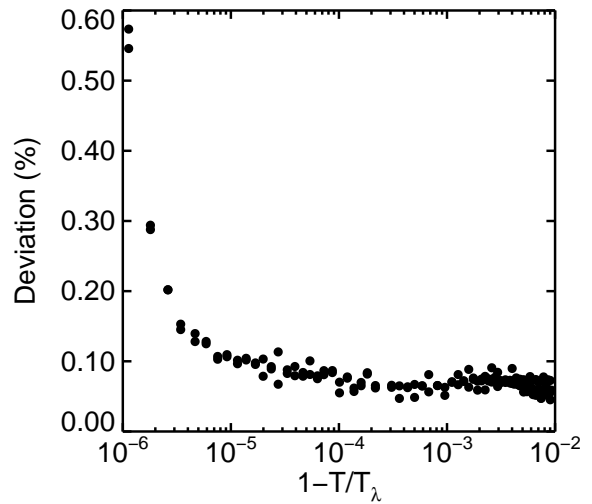}
\caption{\label{fig:epsart8}Fractional deviations of ground specific heat results
from the zero-gravity function.}
\end{figure}
%****************************************************************

\section{\label{sec:level4}FLIGHT OBSERVATIONS \protect\\}

\subsection{\label{sec:level4a}Measurement Sequence}
In this section we summarize the main events of the experiment,
starting with the final refill of the cryostat on the launch pad at the Kennedy Space Flight Center.  
Because the duration of the experiment was limited to 10 days the flight 
measurements were performed in
a sequence designed to obtain the results of most value as quickly as possible.  A conflicting
requirement was to re-calibrate the instrument to verify its performance and provide links to
other specific heat measurements.  As a compromise a limited calibration was performed early in
the sequence as the calorimeter was warmed towards the transition for the first time.  In this
period the HRTs were re-calibrated against the GRTs and the power dependence of the heat capacity
results was measured. The experiment was run automatically by its computer until the second day
of high-resolution measurements. Manual control was then established to optimize the data
collection by allowing for the effects of cosmic rays and heating from particles trapped in the
magnetosphere over the South Atlantic ocean. The remainder 
of the flight period was devoted to heat capacity measurements near the transition.

Shortly before launch the instrument was cooled from 4.2 K to about 1.7 K, the 
cryostat was refilled with superfluid and $^{3}$He exchange gas was placed in the instrument
vacuum space to provide a cooling path to the bath for heat dissipation in the instrument during launch.
About two hours after launch the experiment was switched on and the calorimeter temperature was
found to be close to 2.02 K.  A motor-driven valve on the instrument evacuation line was opened
and the charcoal adsorption pump in the vacuum can was warmed to $\sim$ 40 K to purge it of $^{3}$He.
When the pressure in the vacuum can fell below an estimated 100 microns the valve was closed and
the adsorption pump was cooled. The cryostat was also warmed to its maximum safe operating
temperature in space, close to 2.1 K, to aid in releasing adsorbed $^{3}$He 
from the vacuum can walls. While
the cryopump was reducing the pressure in the instrument, some preliminary tasks in the
measurement sequence were performed.  The first step involved warming the SQUIDs above their
superconducting transition temperature ($\sim 9$ K) and cooling them slowly to 4.5 K.  
This reduced any flux
trapped in the SQUID junctions and in the shielding tubes, improving performance. The next event
of significance was the spontaneous generation of a small helium bubble in the calorimeter which
established a thermodynamic path at the vapor pressure for the heat capacity
measurements.  Shortly after the bubble was formed, the cryostat venting rate was increased, 
allowing it to
cool to its minimum operating temperature of $\sim 1.7$ K. This step allowed the calorimeter to be
brought under thermal control and also produced significant cryopumping of the residual $^{3}$He gas
on the vacuum can walls.

The next step was to prepare the instrument for high resolution measurements. To do this, the various 
stages of the thermal isolation system were brought to their normal operating temperatures which 
were kept as close as practical to that of the calorimeter. The heaters on all the HRT pickup 
loops were then activated to null the circulating persistent currents.  Next, the HRT calibration 
procedure was performed between 2.126 and 2.156 K. The stage 4 servo was then set to the fine 
control mode and the drift rate of the calorimeter temperature was reduced to a value acceptable 
for heat capacity measurements.  It was possible to reduce this drift rate to $\sim 10^{-12}$ K/s with 
little difficulty by varying the stage 4 temperature, however, 
variations of heating from the charged particle 
flux made it difficult to maintain this level of stability for more than a few minutes. When 
the pressure in the vacuum can approached the $10^{-9}$ torr range the calibration sequence was
continued.  At this point the power dependence of the heat capacity results was measured.  This step
also verified that the heater driver circuit was operating normally.

After the heater circuit test was completed, the primary heat
capacity measurements were commenced. Since there was insufficient time in the mission to collect more
than one wide range data set, the final 20 mK of the HRT calibrations was interwoven
with the heat capacity measurements.
The GRTs on the calorimeter were deactivated about 1 mK
below T$_{\lambda}$ to reduce temperature gradients in the calorimeter.
The heater on one of the calorimeter HRT pickup
loops was again activated at 709 $\mu$K below the transition, setting the current in the loop to zero once
more.  This was done to minimize a small magnetic coupling effect between the isolation stages 
described earlier.
While the pickup coil leads were in
the non-superconducting state, a small quantity of flux was not measured by the SQUID. This
discontinuity was corrected for by noting the flux change
on the second calorimeter HRT during the time the
heater was on. At 13 $\mu$K below the transition the current in the pickup loop of the second HRT
was nulled. Again, the flux discontinuity was corrected by using the readout of the alternate HRT.

When the calorimeter was 92 nK below the transition, the high resolution
measurement mode was activated.  In this mode the pulse size estimator was disabled and a
set of pulses of constant energy was applied to the calorimeter heater.  This set was designed to
warm the calorimeter through the transition, terminating about 10 nK on the high temperature
side.  Another routine was then used to cool the calorimeter by about 50 nK.  After repeating this
cycle ten times the automatic operating mode was terminated and the instrument was operated from the
ground.  Typical measurement sequences in this phase consisted of sets of 5 to 10 pulses of
1 to 10 nK size starting 10 to 30 nK below the transition.  Each set was followed by a cooling ramp at
a rate of $\sim$ 100 pK/sec to the starting temperature for the next set.  The
location of the lambda point on the HRT flux scale was found to within $\pm 2$ nK
by observation of the transient
overshoot of the calorimeter shell after a pulse, due to the finite thermal conductivity of
the helium above the transition.  In later analysis more precise determinations were made by
modeling the behavior of the cooling rate as the calorimeter passed through the transition. After the
region near the transition was well covered, heat capacity measurements were
extended a few $\mu$K on the high
temperature side. In this region estimates of the thermal conductivity can be obtained from the
relaxation data.  These measurements were taken on a time available basis and were not expected to
approach the accuracy of the results on the low temperature side. After about 10 days of
operation the cryogen was exhausted and the experiment was terminated. The details of the significant
portions of the measurement sequence are given below.

\subsection{\label{sec:level4b}Bubble Detection}
An important aspect of the experiment was to
ensure that the measurements of heat capacity were performed along a thermodynamic path that is as close
as possible to constant pressure.  This was done by allowing a small bubble of vapor to form in the
calorimeter, taking advantage of the slightly higher density of liquid helium in the region
very close to the lambda point. As the sample of helium is warmed from the starting temperature near 2 K
towards the lambda point, the liquid shrinks slightly, generating a bubble.  This event occurs
spontaneously due to the metastability of the thermodynamic state, as opposed to the equilibrium
collapse behavior observed at T$_{B}$ on cooling. \cite{NissenLipaCzech1996}
Nucleation of the bubble can be triggered by an imperfection
on the calorimeter wall or by the passage of a charged particle through the fluid.
In ground testing the resulting
sharp change of the heating rate of the helium was found to occur within 1900 seconds of entering the
metastable region. In flight the event occurred about 1863 seconds after entering the region, but at
2.079 K vs. 2.076 K on the ground. In zero gravity the states occupied by the sample on the helium
phase diagram are slightly different to those on earth. On average, this effect would cause the bubble
to be formed at a temperature about 20 $\mu$K warmer in space, all else being equal. From these
observations we conclude that near 2 K the probability of nucleating a stable bubble in superfluid
helium by a cosmic ray must be $<$ 10$^{-4}$ for pressures on the order of 2000 Pa
below the vapor pressure. 
The temperature of the bubble event can be used to set limits on the amount of helium in the calorimeter. 
Since the bubble event occurred at a slightly warmer temperature in flight, this was an indication 
that no helium had been lost from the calorimeter and the pre-flight mole number was still valid.

\subsection{\label{sec:level4c}Flight Calibration of HRTs}
As noted earlier, comparison of the resistance values at the lambda
point of the two GRTs on the calorimeter over the whole experiment period indicated that GRT $\#$1 was
slightly more stable than $\#$2. For GRT $\#$1 no offsets greater than 30 $\mu$K were detected over the 8
months prior to flight, essentially the limit of our measurement capability. It was also found that
the noise on bridge $\#$1 continued to be lower than that on $\#$2. Due to the constraints of the
mission the HRTs were calibrated over the range 2.126 to 2.176 K,
half that of the ground calibration. A set
of 76 data points spaced at $<$ 1 mK intervals was used to calculate the coefficients of the best fit
function in Eq.(5). The measurements were analyzed in a similar way to the ground calibration data.
A study of ground measurements showed that the smaller range could lead to some bias in the
parameter values obtained from the curve fit. We also found that the value of the parameter
$a$ in Eq.(5) was extremely stable from calibration to calibration. We therefore decided to fix it at
the ground value in the analysis of the flight data and only
fit the other two parameters. This approach was
shown to significantly reduce the bias. In Fig. 9 we plot the resulting differences between the
GRT $\#$1 temperatures and the HRT temperatures derived from Eq.(5), as a function of GRT temperature.
The noise in the plot is almost entirely due to the GRT measurements. It can be seen that there is
little systematic deviation from the function over the entire range of the data. The relative
deviations of the two HRTs over the range used in the heat
capacity measurements were within $\pm35$ nK, and the maximum slope
difference was $<$ 5 parts in $10^{6}$. Close to the
transition the temperature offsets were reduced 
to $< 5\times10^{-10}$ K when estimates of the lambda transition
temperature became available. Comparison with the ground calibrations showed only very small
changes, $< 0.045\%$ in d$\phi$/dT at the lambda point. Also, the
apparent value of T$_\lambda$ on the GRT $\#$1
temperature scale changed by $<$ 3 $\mu$K, giving us confidence that there was little change of the
absolute calibrations in flight.

%****************************************************************
\begin{figure}
\includegraphics{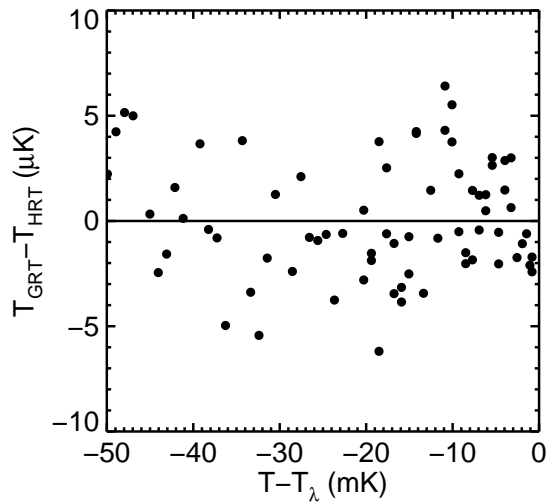}
\caption{\label{fig:epsart9}Flight calibration data for GRT $\#$1
on the calorimeter vs. the HRT scale.}
\end{figure}
%****************************************************************

Since it was not possible to obtain sufficient data to directly
calibrate the HRTs with the persistent current nulls made within 1 mK of the transition, a
cross-calibration technique was used. By activating only one heater at a time and spacing the nulling
temperatures by $\sim$ 700 $\mu$K it was possible to piece together the changes
in the values of d$\phi$/dT to high
accuracy.  For the first section a third order polynomial was
used and for the second, shorter region a second
order polynomial was used. This is all that is needed for the close-in heat capacity data since
curvature effects from the calibration in this region are small. We obtained an uncertainty in the
values of d$\phi$/dT  relative to those from the wide range
calibration of $<$ 0.0075$\%$, which is negligible
compared with the noise of the heat capacity data.

\subsection{\label{sec:level4d}HRT Noise}
The noise on the HRTs was significantly larger in flight than on
the ground.  The excess noise appeared to be related to the charged particle flux passing through the
HRTs.  Protons and alpha particles passed through the HRTs on the order of once per second.  Due to
the 1 second time constant of the thermal link between a HRT and the calorimeter, the
individual hits could not be resolved, but were smeared out and appeared as noise.
Heavier nuclei passed through the HRTs on average every 100 seconds or so.
These generated heating spikes of up to
several nK depending on the length of the track through the sensor assembly. An example of the
HRT output obtained over a few hundred seconds is shown by the upper curve in Fig. 10. The
lower curve shows ground data for comparison. Apart from the large spikes in the data,
the effective noise was about 5 to 8 times the ground value.  In addition, the spike events were
almost always heating, leading to some bias in the averaged signals.
At times of high particle
flux, even higher noise than that shown was encountered. 
A detailed model of the effect of cosmic rays on the
thermometer was developed and the results showed very similar behavior to that seen in the figure. This
model also showed that the average heating caused a temperature offset between the thermometer and the
helium sample of $\sim$ 1-2 nK. The measurement of this offset is described below.
Additional noise events were seen 
which correlated with large acceleration disturbances on the shuttle.
We suspect that this effect was responsible
for the occasional events that indicated apparent transient cooling of the HRTs. Special curve fitting 
techniques were used to minimize the effect of the spikes on the heat capacity results as described below.
%****************************************************************
\begin{figure}
\includegraphics{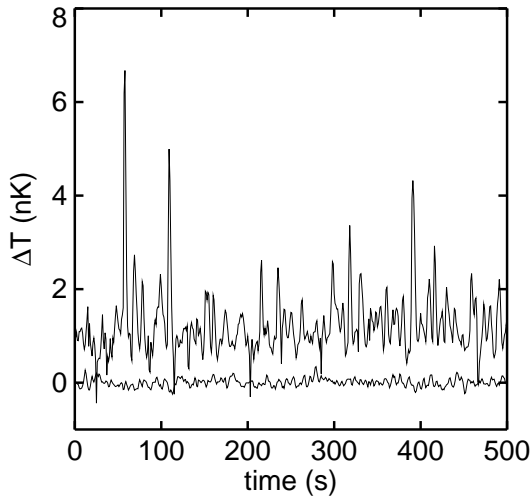}
\caption{\label{fig:epsart10} HRT noise measurements. Upper curve: noise seen
during the mission. Lower curve: similar data collected during ground testing.}
\end{figure}
%****************************************************************

\subsection{\label{sec:level4e}Heat Capacity Measurements}
The initial flight heat capacity measurements were generated with
the same software routines as used for the ground data. The first
group of heat capacity measurements
was a check on their apparent power dependence. The pulses were
applied when the calorimeter was
at 2.1548 K and were similar to those used in pre-flight testing.
The results were more
noisy than on the ground, but otherwise showed no significant change.
The apparent stray power was
$-9.77\times10^{-11}$ W. 
For comparison, the minimum power used in the flight measurements
was $1.7\times10^{-8}$ W. Also, the high power
measurements showed a self-heating effect similar to that on the ground.
Corrections were
again applied for these effects. The second group of data
consisted of wide range measurements performed in
a similar manner to the ground. The other two groups of data
were high resolution measurements
spanning the region not accessible on the ground, and measurements
above the transition. A set
of high resolution measurements passing through the
transition is shown in Fig. 11, with
individual temperature steps of $\sim 4$ nK. 
The growth of the
relaxation time as the sample moves deeper into the normal phase is
apparent.  Measurements were made with step heights as small
as 1 nK, but at this resolution the
results were extremely noisy, with typical uncertainties of $\sim 30\%$.
The practical limit to the resolution
of the experiment from the point of view of heat capacity measurements
was a step height of about 2 nK.  Even here, a
large amount of averaging was necessary to produce a usable result.

%****************************************************************
\begin{figure}
\includegraphics{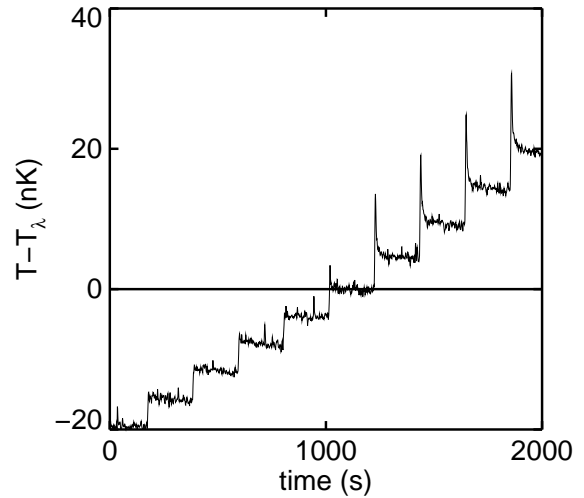}
\caption{\label{fig:epsart11}High resolution temperature vs. time data containing a number
of heat pulses with step heights of $\sim 4$ nK.  The estimated location of the
lambda transition is marked.}
\end{figure}
%****************************************************************
During the experiment 876 heat pulses were applied giving rise to 1752 temperature
step measurements and heat capacity values. Approximately 3$\%$ of the measurements were
abandoned due to excessive radiation heating, faulty telemetry
records, or large external disturbances.
Twenty-nine data points were deleted on the basis of being more
than $5 \sigma$ from the best fit function.
A number of data sets were timed to take advantage of the
most favorable conditions: periods of low and stable radiation levels, quiet times on the shuttle and
highest vacuum in the instrument. Some measurements
were also performed after the altitude of the shuttle
was lowered from 288 to 210 km. This period was associated with a somewhat improved noise level in the
HRTs but was unfortunately too brief to affect the overall accuracy of the experiment
significantly. The conversion of the data to specific heat values is described in section VI.

\subsection{\label{sec:level4f}Lambda Point Location}
As part of the process of generating the heat capacity data it was
necessary to repeatedly cool the calorimeter back to a starting temperature 
for the next pulse sequence.
A cooling ramp was initiated by reducing T$_{4}$ sufficiently to cool the calorimeter at $\sim$ 0.1
nK/sec. As the sample cooled through the transition the outer layer of normal helium nearest the
calorimeter wall converted to the superfluid state. The boundary between 
the two phases then propagated to
the center of the sample in a time span of several seconds. While the helium was in this two-phase
region, T$_{5}$ paused very near T$_{\lambda}$ until the temperature gradient
within the helium was eliminated. When
all of the helium was transformed into He-II, the sample cooled again, but at a slower
rate than above the transition, due to the higher heat capacity in this region. 
This signature was
very useful for keeping track of the apparent location of the lambda point on the flux scales of the
HRTs. A typical cooling ramp is shown in Fig. 12.  Also shown by the smooth curve 
is a model of the behavior fitted to the
average cooling rate and
adjusted to give the best estimate of T$_{\lambda}$. A few heating ramps were also
performed, but these had a less clear transition signal due to the absence of the pause at the
normal/superfluid phase boundary.

%****************************************************************
\begin{figure}
\includegraphics{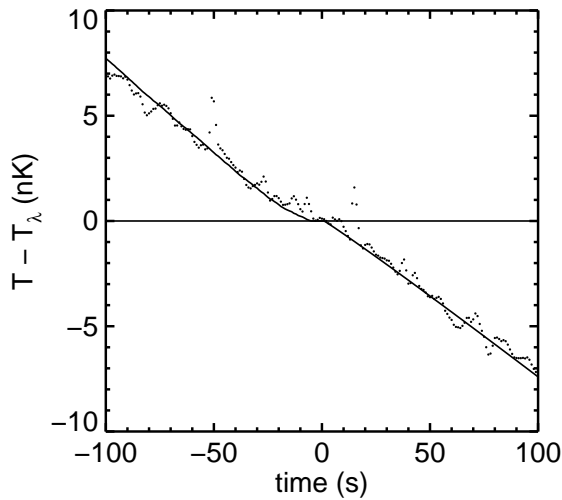}
\caption{\label{fig:epsart12}An example of a cooling curve through the lambda point.  Solid
line is the result of a model calculation. The location of the lambda transition
is determined by fitting the model to the data.}
\end{figure}
%****************************************************************
Some corrections need to be applied to the apparent lambda point
flux readings to obtain the best estimates suitable for use with the heat capacity data. The major
correction was due to the residual gas in the instrument vacuum space. This gas introduced a small
temperature gradient in the calorimeter and HRT assembly related to
the offset T$_{5}$ - T$_{4}$ and to P$_{4}$.  This
was large enough that the variation of pressure with time introduced a detectable shift in the
apparent lambda-transition temperature.  It was therefore necessary to understand this effect well
enough to correct the HRT temperature scales at times intermediate between lambda-point
determinations.  In addition to the general trend of pressure with time,
there were two phenomena that helped
us calibrate the effect. First, on the sixth day of the mission, a heater in the main helium bath
was turned on briefly in an attempt to measure the amount of cryogen remaining. This event warmed the
instrument walls, releasing some gas trapped there and causing a pressure increase lasting a
number of hours. This resulted in a substantial change in the lambda-point flux readings over the
same period. Second, we observed a rapid thermal transient in T$_{5}$ when P$_{4}$
was changed.  This data allowed us to determine two pairs of parameters which appeared to be the
most significant in a model of the gas effect.  These parameters were the coefficients of the coupling
between the HRT flux readings and T$_{4}$, and P$_{4}$ at constant T$_{4}$.
We found that the coupling to P$_{4}$
was the more important of the two effects.  
For the lowest gas pressure in the mission, the
coefficients for HRT $\#1$ were 29.4 mK/W for the heater coefficient, and 0.11 $\mu$K/K for
the temperature
coefficient. For HRT $\#2$ we obtained 19.7 mK/W and 0.21 $\mu$K/K respectively. 
At the start of the high
resolution portion of the mission, the temperature coefficients were about a factor of two higher,
due to the higher gas pressure. Fig. 13(a) shows the apparent location of the lambda point as a
function of time for HRT $\#1$. Fig. 13(b) shows the residuals after removing the modeled effects. The
r.m.s. uncertainty of the corrected locations is $\sim 0.3$ nK. Given the quality of the fit we
estimate an uncertainty in the lambda temperature at any time after the start of the high resolution
measurements of $\pm 0.5$ nK.

%****************************************************************
\begin{figure}
\includegraphics{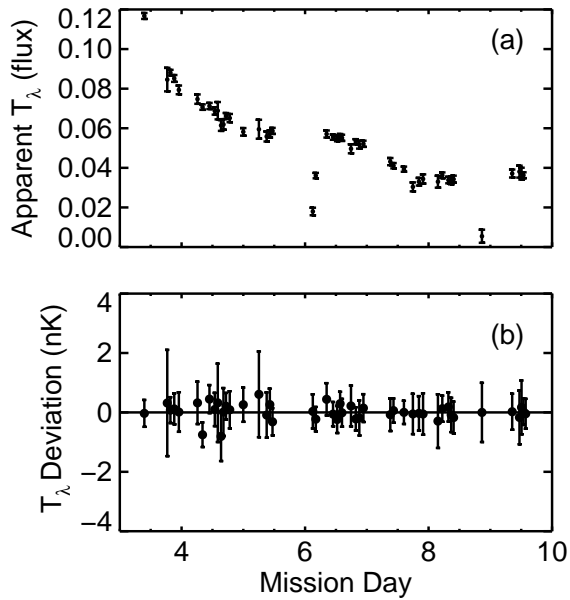}
\caption{\label{fig:epsart13}a) Time dependence of the apparent HRT flux readings
corresponding to the lambda transition during the high resolution phase of the mission.
b) Residuals for the lambda point temperatures
vs time after correction for pressure
and charged particle heating effects.}
\end{figure}
%****************************************************************

\subsection{\label{sec:level4g}Charged Particle Data}
It was observed that the heating rate of the calorimeter fluctuated
slightly, typically with twice orbital period, preventing the calorimeter from reaching the level of
thermal stability achieved on the ground.  This variability was
correlated with the readout of the CPMs. It
can be seen in Fig. 14 which shows in the lower trace the temperature
of the calorimeter over a 5000-second period
where no pulses were applied.
The upper trace
shows the output from one of the CPM's over the same period 
averaged over 200 s intervals, clearly
demonstrating the correlation.
Correction factors were obtained by fitting the HRT variation
vs. charged particle flux data with a 3-parameter function. The first parameter
described the heat leak to stage 4 which is constant over the period of the data. The second
parameter was proportional to the particle flux and was a
measure of the heat input to the calorimeter due
to cosmic rays. The third parameter accounted for the temperature offset between the HRT and the
calorimeter due to cosmic ray energy dissipation in the HRT. We obtained temperature offsets for the
two HRTs of 0.45 and 0.65 nK/count/sec, and a heat input to 
the calorimeter of $1.29\times10^{-10}$ J/count,
where the count number was derived from a model of the CPM responses to the known cosmic ray energy
spectrum. \cite{WilliamsonLipa1995Cosmic} These correction factors were
applied to all the high resolution thermometry data
and the power input coefficient was used to correct the drift rate during heat capacity
measurements. It was also found that for several hundred seconds after 
a passage through the radiation zone over the South Atlantic
the heat capacity data appeared to be slightly affected.
Data from this region were not used. A more detailed account of the charged particle
observations is given elsewhere. \cite{WilliamsonLipa1995Cosmic}

%****************************************************************
\begin{figure}
\includegraphics{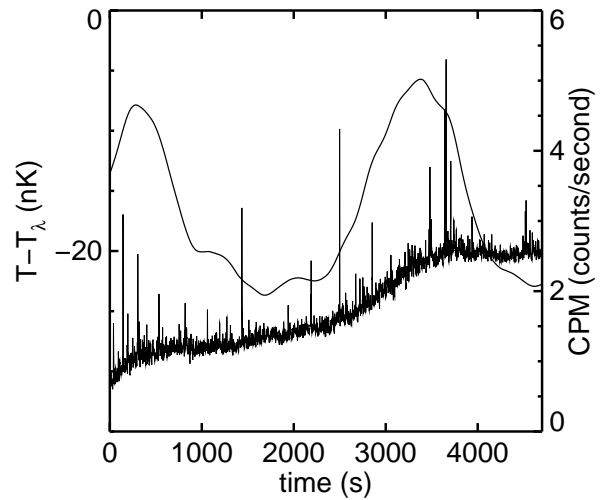}
\caption{\label{fig:epsart14}Input data for calibrating the cosmic ray heating.  Upper line:
smoothed CPM data (right scale); lower line: raw HRT signal showing variation of calorimeter
heating with time (left scale).}
\end{figure}
%****************************************************************

\subsection{\label{sec:level4h}Accelerometer Data}
A significant amount of accelerometer data was available during the
mission. In addition to the peak acceleration in each 1-second interval, power spectral density plots
were available at approximately 100 min intervals and on special request.  It was found that
the peak acceleration amplitude was typically in the range
$0.5 - 1\times10^{-3}$ g with spikes to $\sim3\times10^{-3}$ g occurring
approximately every few hundred seconds. From the larger acceleration events we were able
to determine some characteristics of the acceleration sensitivity of the instrument.  We
found that the HRTs had a sensitivity to steady state acceleration, possibly due to changes of
mechanical stress or some form of magnetic coupling.  This was evident from HRT 
data collected during low
power operation of the shuttle engines which applied a transverse acceleration to the instrument.
The sensitivities of the two HRTs to this acceleration were 1.1 and 0.8 $\mu$K/g. 
This indicates that some
of the larger acceleration transients during normal operations should be visible on the HRTs.  We
searched for noise spike correlations with some of the largest events seen on the accelerometer
output. A significant number of correlations were found, indicating that up to a few percent of the
spikes in the HRT data were due to acceleration transients.  However since there were clearly many more
cases with no correlation, in the data analysis we elected to treat the acceleration spikes in the same
manner as charged particle events. It was also apparent that most of the acceleration-related
transients on the HRTs corresponded to heating.

The HRTs also showed a noise increase of about a factor of two during the
low power operation of the shuttle engines. Since the acceleration noise increased by
roughly a factor of eight during this period, it is unlikely that vibration contributed significantly
to the noise level seen during normal operations. An apparent power
input to the calorimeter averaging $\sim2\times10^{-9}$ W was also
seen in this period. Linear extrapolation would indicate that levels
of $2-3\times10^{-10}$ W might occur during normal heat capacity
measurements, which is somewhat less than the measured
charged particle dissipation level. However, this estimate is extremely crude, given the
mechanical complexities involved. Our experience with low level vibration inputs on the ground
indicated a highly nonlinear situation. The flight spectral information indicated highly variable
vibration levels in the 1 Hz band centered on 55 Hz, and
the corresponding heating estimate was approximately
consistent with the above value.
Further details of acceleration effects are available. \cite{NissenLipaAIAA}

\section{\label{sec:level5}POST-FLIGHT MEASUREMENTS \protect\\} 
After the instrument was returned to JPL it was operated in its
flight configuration once more and a number of tests were performed to verify the stability of various
parameters. Since the burst discs on the calorimeter were designed to rupture at the end of the flight,
only tests with an empty calorimeter could be performed.  Also since the cryostat had reached room
temperature, the magnetic flux trapped in the HRTs was dissipated.
The post-flight test activities were designed 
to improve the validation of the science data by showing that
the calibration of various A/D converters 
and electronic circuits had not changed significantly from the pre-launch calibration.

Initially the electronics were operated alone using dummy loads to
simulate the instrument. First, the calibrations of the bridges for
the GRTs on the calorimeter were
checked using a reference decade resistance box. 
It was found that the balance points of bridges $\#1$
and 2 had changed by $- 0.7 \pm 0.1$ $\Omega$ and $+ 0.8 \pm 0.1$ $\Omega$ over the range used.
These shifts were 
small enough to be neglected. Next, the heater pulse shape was examined
for changes in the transient behavior.
No changes exceeding 10 $\mu$s duration were detected.
Also, the pulse timing was checked and found to be accurate to $\pm$0.0003$\%$.
The calibration of the A/D converter in
the heater current monitoring circuit was checked.  The results agreed with the pre-flight
measurements to within $0.02\%$ in power. Similar results verified correct operation for all the heater power
ranges. 

The electronics were reconnected to the instrument and the cryostat
was cooled to $\sim 2$ K.
The calibrations of the A/D converters monitoring the
calorimeter HRTs in terms of bits/$\Phi_{0}$ were
checked and found to be unchanged from flight to within $0.004\%$. This indicates that the
converters and the SQUID feedback electronics were probably working normally during the flight. The
temperature and power dependence of the resistance of the heater on the calorimeter were also
remeasured.  A zero-power resistance offset of $0.022\%$ was found. The power dependence of 
the heater resistance was found to be about an order of
magnitude higher than observed in the flight. It is conjectured that some damage occurred to the
heater bonding when the calorimeter warmed up after the mission. 
In the analysis below, the pre-flight value of the heater resistance was used.

A 100 Gauss field was then
trapped in the HRT flux tubes and their sensitivities at the lambda point
were verified to within $2\%$.
During the following few years, the electronics system
was used on a second flight experiment \cite{LipaPhyRevLett2000} and no
significant anomalies were detected.

\section{\label{sec:level6}DATA ANALYSIS \protect\\}
In the ideal case the analysis of heat capacity data collected by
the pulse method is straightforward.  However, when high resolution measurements are attempted
under adverse conditions a number of error sources need to be considered. The analysis of
the raw data from the ground measurements was described earlier. Additional errors arise from
operation in the space environment. In this section we describe the details of the flight specific
heat calculations and our models of the associated corrections. We also describe the results of
the curve fitting analysis.

\subsection{\label{sec:level6a}Specific Heat Calculation}
The T$_{5}$ drift data before and after each pulse were corrected for the
heating of the calorimeter by cosmic rays.  This correction term was determined from the CPM outputs
as described earlier. These corrections were applied so that the temperature at the center of the
heat pulse was undisturbed, allowing an accurate correspondence between the calculated
midpoint temperature of a pulse and the original HRT temperature scale. The line fitting routine
used to determine the intercepts was chosen to reduce the sensitivity of the results to the
non-Gaussian spikes in the data. \cite{NumericalRecipes1987}  The routine was 
also utilized to identify and remove the larger
cosmic ray events from the data set. Spike events extending
more than $2.5 \sigma$ from the fitted line were
identified and removed, and a second fit performed. Model testing showed that these
precautions reduced the bias in the heat capacity results substantially.

%****************************************************************
\begin{figure}
\includegraphics{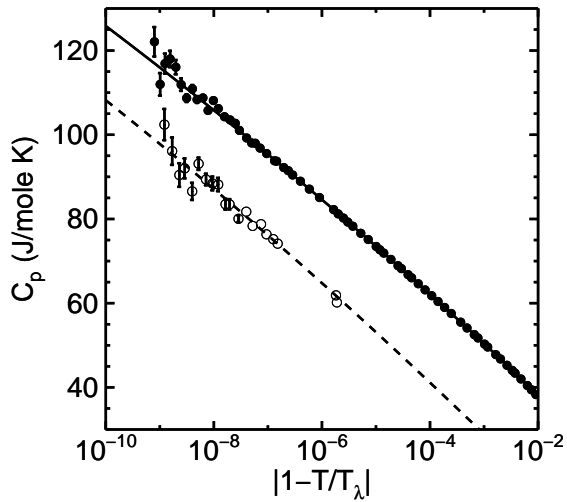}
\caption{\label{fig:epsart15}Semi-logarithmic plot of the specific heat vs. 
reduced temperature over the full range measured. Below the transition the 
data (closed symbols) were binned with a density of 10 bins per decade, 
and above (open symbols) with a density of 8 bins per decade.
Lines show best fits to the data.}
\end{figure}
%****************************************************************

The statistical weight of the input temperature data, which is normally
inversely proportional to the thermometer noise squared, was modified slightly to allow for the
uncertainties in the heating rate corrections and the offset of the HRT temperature from the calorimeter
temperature. The corrections had the effect of slightly deweighting the data at higher
count rates as well as de- weighting the data further from the center of the heat pulse where the
correction becomes larger. The uncertainty in the heat capacity was derived from the usual partial
derivatives and the uncertainties reported from the least squares fit. From a model of the
effect of noise on the results we found that the heat
capacity determined from the average of many noisy
pulses, $\langle\text{C}_{p}\rangle$, was slightly greater than that
determined from a noise-free pulse. 
It is easy to show that
$\langle\text{C}_{p}\rangle/\text{C}_{p} = 1 + (\sigma_{\text{T}}/\Delta\text{T})^{2} + ...$.
This correction term becomes important for very small step heights
where averaging is used to improve the
signal/noise ratio. All data were corrected for this effect
using the statistical $\sigma_{\text{T}}$ for the individual
measurements. The curvature correction described earlier was also 
applied to the results, along with the path corrections given by Eqs. 6 and 7.

We found that
we could decrease the scatter of the results by performing a more complex
fit over all the pulses in a single data set
simultaneously.  This was accomplished by calculating the heat leak to stage
4 over many (N $\sim 10$) pulses and 
performing a least squares fit with N temperature steps,
one at each heat pulse.  This reduces the total
number of parameters determined for the N pulses: there is now only one straight line fit and N step 
heights compared to N line fits and N steps if each pulse is treated separately. This essentially takes 
advantage of the very long relaxation time for T$_{5}$ to T$_{4}$, and assumes
that the main perturbations to 
T$_{5}$ have been modeled correctly. The specific heat results
are shown in Fig. 15 after being bin-averaged 
at a density of 10 bins per decade of $t$ below the 
transition, and 8 bins per decade above the transition. It can easily be seen that the
specific heat divergence continues to the 
highest resolution we were able to achieve.

%****************************************************************
\begin{figure}
\includegraphics{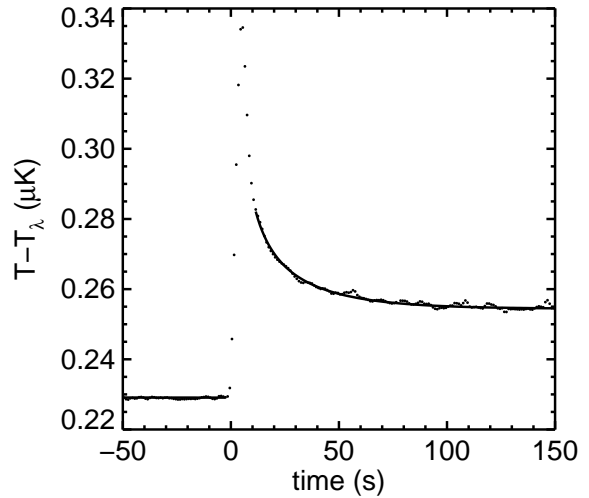}
\caption{\label{fig:epsart16}Example of a HRT response to a heat pulse well above the lambda
point. The lines are the results of a least squares fit to the data on each side of the pulse.}
\end{figure}
%****************************************************************

\subsection{\label{sec:level6b}High Temperature Data}
Analysis of the data above T$_{\lambda}$ was complicated by the finite thermal
conductivity of the normal helium. This introduces a transient immediately after a heat pulse
consisting of an overshoot in the temperature of the calorimeter
shell and a rapid decay
as the heat in the shell diffuses into the helium. An example of a pulse above the transition is
shown in Fig. 16. The extra parameters required to allow for this effect increase the scatter of the
heat capacity results.
Furthermore, as the main objective of the experiment was to measure C$_{p}$
below T$_{\lambda}$, only a 
few measurements were made above T$_{\lambda}$.
The net result was a superior
signal-to-noise ratio in the results below T$_{\lambda}$.

The thermal behavior of the shell was modeled using a radial
diffusive heat flow approximation to the power input from the heater. This model was justified
based on the very high conductivity of the calorimeter shell relative to the helium above the
transition, the absence of convection in space and the high degree of isolation from the surroundings.
An analytic solution to the problem of a transient heat input to the surface of an isolated solid
sphere with diffusive heat flow is available. \cite{OzisikBVPHeatCond1989} 
The temperature of the surface as a function of
time after the pulse is given by:
%=============================================================
\begin{equation}
\text{T}_{5} = \text{T}_{f}+ \sum_{n=1}^{\infty}C_{n}\text{ exp}(\text{-t}/\tau_{n}) 
\end{equation} 
%============================================================ 
where $C_{n}$ are known coefficients determined by the initial conditions,  
$\tau_{n}= C_{p}/(v\kappa\beta_{n}^{2})$, $v$ is the molar volume of the liquid,
$\kappa$ is its thermal conductivity, 
$\beta_{n}$ are the positive roots of: $\beta\text{a ctn}(\beta\text{a}) = 1$, 
a is the radius of the sample
and T$_{f}$ is the final temperature. We used  the first nine terms of this
series to model the observed relaxation behavior.
The ratios of the coefficients were obtained by modeling the transient
heating of the calorimeter during the pulse
and using the resulting
temperature distribution as the initial condition for the solution
to the thermal relaxation behavior of an
isolated sphere. To allow for the variation of the properties of helium during the decay, the
diffusivity, $v\kappa/C_{p}$, was modeled as having a locally linear dependence on the temperature
within the sample. An example of a fit is shown by the solid line
in Fig. 16. It can be seen that a reasonable
representation of the behavior is obtained over a significant portion of the decay. The data above the
transition became progressively more difficult to analyze as 
the temperature was increased.  This was due
primarily to the increased length of the extrapolation back to the center of the pulse after the
thermal transient had decayed sufficiently. The bin averaged specific heat results 
near the transition are shown on a linear scale in Fig. 17.

%****************************************************************
\begin{figure}
\includegraphics{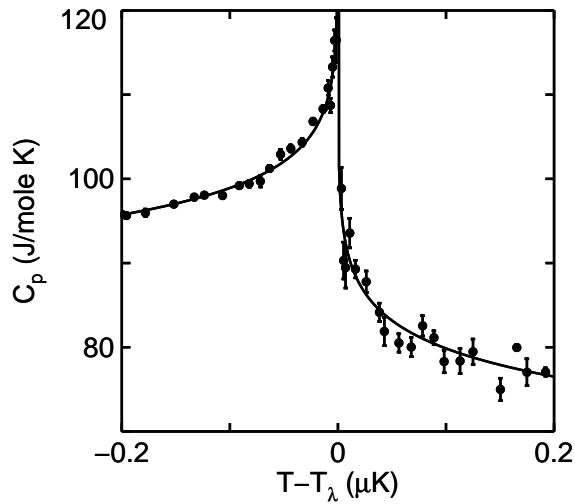}
\caption{\label{fig:epsart17}Bin-averaged data close to the transition. Line shows the best
fit function.}
\end{figure}
%****************************************************************

\subsection{\label{sec:level6c}Curve Fitting}
As described in the introduction, the RG theory makes a prediction
for the critical exponent $\alpha$, describing the divergence of the
heat capacity near the transition and for
the ratio of the leading order coefficients on the two sides of the transition.  The asymptotic form
for the heat capacity near the transition is expected to be given by Eq.(1). We fit the results over
the whole range measured with the truncated trial function:
%---------------------------------------			
\begin{eqnarray}
\text{C}_{p} = \frac{\text{A}^{-}}{\alpha}t^{-\alpha}
(1+ a^{-}_{c}t^{\Delta}
+ b^{-}_{c}t^{2\Delta}) + \text{B}^{-}~~, \text{T}<\text{T}_{\lambda}\nonumber\\
= \frac{\text{A}^{+}}{\alpha}|t|^{-\alpha}
+ \text{B}^{-}~~,~~~~~~~~~~~~~~~~~~~~~~~~\text{T}>\text{T}_{\lambda} \label{8}
\end{eqnarray} 
%---------------------------------------
where we have assumed the constraint B$^{+}$ = B$^{-}$. The simpler form
was used above T$_\lambda$ because the data extend only
to $|t| \sim10^{-6}$, where the additional terms would
still be negligible. All parameters were allowed to vary except
for $\Delta$, which was fixed at its
theoretical value \cite{GuidaJPhysA1998} of 0.529, and T$_\lambda$,
which was determined as described earlier.
See EPAPS \cite{epapsdata} for the complete set of raw data used in the curve fitting.
Also listed is the bin-averaged data set shown in Fig 15.

The best fit values for the parameters are
listed in the first line of Table II along with the ratio $\text{A}^{+}/\text{A}^{-}$.
The corresponding uncertainties are listed
below the values and refer to the standard statistical error
evaluated from the curve fitting routine.
The uncertainties for the derived quantities A+/A-
and P were evaluated by the usual formulae for propagation of errors \cite{Bevington}
taking into account the strong correlation between the fitted parameters
$\alpha$, A+ and A-.
To obtain some feel for the sensitivity of the results 
to small changes in the analysis, we performed a
number of extra fits to the data. The second group in the table shows the effect of modifying Eq.(9)
to the form:
%---------------------------------------
\begin{eqnarray}
\text{C}_{p} = \frac{\text{A}^{-}}{\alpha}t^{-\alpha}
(1+ a^{-}_{c}t^{\Delta})
+ b^{-}_{c}t + \text{B}^{-}~~, \text{T}<\text{T}_{\lambda}\nonumber\\
= \frac{\text{A}^{+}}{\alpha}|t|^{-\alpha}
+ \text{B}^{-}~~,~~~~~~~~~~~~~~~~~~~~~\text{T}>\text{T}_{\lambda} \label{9}
\end{eqnarray} 
%---------------------------------------
which treats the third order term as a regular background contribution. It can be seen 
that the shift of $\alpha$ is slightly outside the combined uncertainties, significantly larger than
expected.
This effect can be traced to the large value obtained for $b^{-}_{c}$ which is much
greater than the value of $c_{1}$ (Eq. 3) estimated from the regular background contribution.
It therefore appears that the second order Wegner term  $b^{-}_{c}t^{2\Delta}$ is
the more important term to include in the fitting function, leading to Eq.(9) as the preferred
representation of the behavior.
This effect may be indicative of a rapid crossover to mean field
behavior far from the transition.
In the third and fourth groups we
investigate the effect of reducing the fitting range at each end. Little effect is seen. The fifth
group shows the sensitivity to a shift of the transition temperature of 1 nK,
about twice the estimated
uncertainty in the individual measurements.
The next two groups show the effect of eliminating either the lowest or the highest
power pulses from the fit. There is little dependence on the low power pulses as might be expected
from their high uncertainties, but the high power pulses are important in the fit. For this
case we also find the uncertainty in $\alpha$ has increased by $75\%$. 
To lessen the dependence of the fit
on the measurements with the greatest statistical weight, we set a
lower limit of $\sigma_{\text{C}}/\text{C} = 0.02\%$ and obtained
the results in the eighth group. This constraint had little effect other than to increase the
uncertainties of the parameters.

%--------------------------------------------------------------------------
\begin{table*}
\caption{\label{tab:table2}Results from curve fitting to the specific heat measurements using Eq.(9)
except where noted. Statistical uncertainties are given in parentheses beneath the values.}
\begin{ruledtabular}
\begin{tabular}{ccccccccc}
 Constraint&$\alpha$&A$^{+}$/A$^{-}$&A$^{-}$&B$^{-}$&$a^{-}_{c}$&$b^{-}_{c}$&$P$&Range of fit\\
\hline

Eq. (9)& -0.01264 & 1.05251  & 5.6537 & 460.19 & -0.0157 & 0.3311   & 4.154& $5\times10^{-10} < t < 10^{-2}$  \\
       &(0.00024) & (0.0011)& (0.015) & (7.3)  & (0.0015)& (0.011) & (0.022)&                            \\

Eq. (10)& -0.01321 & 1.05490 & 5.6950 & 443.76 & -0.0253 & -128.4   & 4.155& $5\times10^{-10} < t < 10^{-2}$ \\
       & (0.00025)& (0.0011)& (0.092) & (7.0) & (0.0015)& (2.5)    &  (0.022)&                            \\

Reduced range&-0.01254 & 1.05210 & 5.6458 & 463.11 & -0.0136 & 0.3035	& 4.154  & $5\times10^{-10} < t < 3\times10^{-3}$ \\
       & (0.00043)&  (0.0018)  &  (0.030) & (13.4) & (0.0043)& (0.044) & (0.022) &                            \\

Reduced range&-0.01264 & 1.05251 & 5.6537   & 460.20 & -0.0157	& 0.3311   & 4.154   & $10^{-9} < t < 10^{-2}$    \\
          & (0.00024) & (0.0011) &  (0.015 ) & (7.4)  & (0.0015)& (0.012)  & (0.022) &                            \\

T$_{\lambda}$ + 1 nK&-0.01278 & 1.05307 & 5.6623   &  455.80  & -0.0165  & 0.3372  & 4.151   & $5\times10^{-10} < t < 10^{-2}$ \\
       &      (0.00024)      & (0.0011) & (0.015) & (7.2)  & (0.0015)   & (0.012)& (0.022) &                            \\

P$_{5}>10^{-7}$ W &-0.01269 & 1.05273 &  5.6570   & 458.55   & -0.0160  & 0.3335 & 4.154 & $5\times10^{-10} < t < 10^{-2}$ \\
         &      (0.00026)  & (0.0012) & (0.017)  & (8.0)    & (0.0017) & (0.013)& (0.025) &                            \\

P$_{5}<5\times10^{-4}$ W &-0.01323 & 1.05498 & 5.6970  & 443.27 &  -0.0228 & 0.3853  & 4.156  & $5\times10^{-10} < t < 10^{-2}$ \\
              &      (0.00042)    & (0.0018)& (0.029) & (11.6)  & (0.0038) & (0.028)&  (0.022) &                            \\

$\sigma > 0.02\%$     & -0.01275 & 1.05297  & 5.6620 & 456.89 & -0.0176  & 0.3473   & 4.154  & $5\times10^{-10} < t < 10^{-2}$ \\
               &       (0.00041) &  (0.0018)  & (0.028) & (12.3 ) & (0.0034) & (0.025 ) & (0.022) &                            \\

\end{tabular}
\end{ruledtabular}
\end{table*}

%--------------------------------------------------------------------------

Taking into account the effects of the various constraints it appears that
the flight results indicate  $\alpha = -0.0127 \pm 0.0003$ and
A$^{+}$/A$^{-} = 1.053 \pm 0.002$ with a high
degree of confidence. These values can be compared with the theoretical estimates given in Table I. It
can be seen that our result for $\alpha$ falls between the two recent estimates, giving us confidence in the
overall correctness of the RG approach. If the discrepancy between the two estimates can be resolved, a
very high quality test of the theory would result.  It is interesting to note that our result falls
very close to the value  -0.01294 $\pm$ 0.0006  obtained earlier
by Kleinert. \cite{KleinertPRevD1999}
This suggests that the earlier method of resumming the perturbation expansion
may be more reliable than the more recent one. \cite{Kleinert2000}

The value we obtain for the ratio A$^{+}$/A$^{-}$ can be
compared with the calculation of Str\"osser,
M\"onnigmann and Dohm \cite{StrosserDohm2000} who's result for
$P$ implies A$^{+}$/A$^{-} = 1.056 \pm 0.003$ if $\alpha = -0.0127$. 
For the value of $P$ itself,
we obtain $P = 4.154\pm 0.022$ where the uncertainty is the $1 \sigma$ statistical
value including the effects of
correlations of the parameters.
This compares well with the Str\"osser et al. result of $4.39\pm 0.26$.
Very recently, Str\"osser and Dohm \cite{StrosserDohmPRevE2003} have obtained
the improved result $P = 4.433\pm 0.077$ from a 4-loop analytic calculation.
On the experimental side, G. Ahlers has informed us \cite{AlhersComm} that his 
data \cite{AhlersPRevA1971} are consistent with $P = 4.194\pm 0.019$, quite
close to the present measurement.
We note that the value of P is dependent on the 
heat capacity behavior above T$_\lambda$ which is not very
well established in our experiment.
A value of $P$ can also be
derived from the data of Lipa and Chui. \cite{LipaPhyRevLett1983}
From their published analysis,
which did not constrain B$^{+}$ = B$^{-}$,
we obtain $P = 4.57 \pm 0.4$.
We note that the value of $P$ and its uncertainty are significantly affected by
the fitting constraint B$^{+}$ = B$^{-}$.
This appears to be a result of the strong correlations between
A$^{+}$ and B$^{+}$ and A$^{-}$ and B$^{-}$ in the 
curve fitting. If this constraint is added to the 
analysis of ref. 5 we obtain $P$ = 3.98 $\pm$ 0.02. 
More work would be valuable to determine an accurate value of P.

The quality of the fit to the model can be seen in Fig. 18 where the
residuals from Eq.(9) are shown below the transition. 
Part (a) shows the bin-averaged results 
over the full range measured, and part (b)
shows the individual
measurements in the range $t > 10^{-7}$. 
The bin density was 10 per decade,
evenly spaced on a logarithmic scale.
The parameters of the reference function are given in line 1 of Table II. 
Little curvature in the
deviations at large $t$ is evident, 
indicating residual effects from the
truncation of the fitting function and 
from the calibration approximations are small.
The deviation plot also shows
that there is no indication of rounding from
gravity or finite size effects near the transition. 
An estimate of the effect of the 
surface specific heat is shown by the broken curve.
This result extends the agreement with the predicted
behavior two orders of magnitude closer to the transition than previously.

%****************************************************************
\begin{figure}
\includegraphics{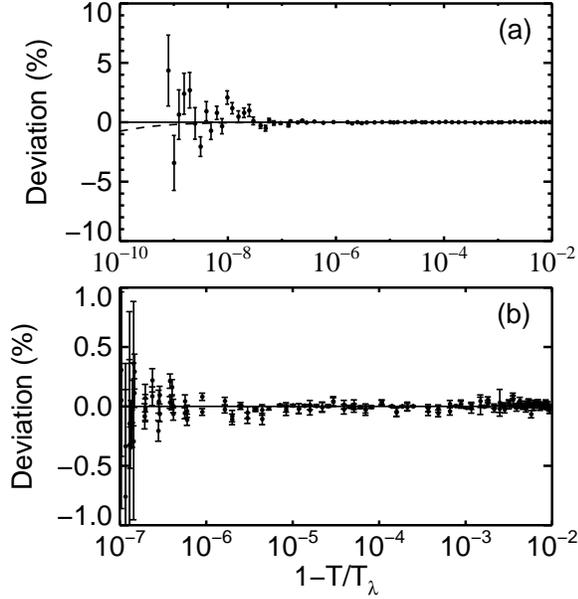}
\caption{\label{fig:epsart18}a: Deviations of the bin averaged flight data from the best fit
function vs reduced temperature on a semi-logarithmic scale below the transition. Dashed
line: estimated surface specific heat term. b: Magnified view of deviations of
individual measurements for $t > 10^{-7}$.}
\end{figure}
%****************************************************************

The results given here for $\alpha$ and A$^{+}$/A$^{-}$ differ somewhat from those quoted in
our preliminary analysis \cite{LipaPhyRevLett1996} after correction
for a computational error. \cite{LipaPhyRevLett2000}
Apart from the addition of the high power pulse data the
main differences between the two analyses are improved modeling of the HRT calibrations 
and the power dependence of the heat capacity.
Additional analysis has shown that approximately $90\%$ of the change in $\alpha$
is due to the improved calibration function in Eq.(5).
Given the small change of $\alpha$ that has
resulted from this additional work, we have 
confidence that systematic effects are now well controlled and
the present results are quite robust relative to 
alternative treatments of the raw measurements.

We can use our result for $\alpha$ to perform a test of the Josephson
scaling relation $3\zeta + \alpha -2 = 0$. 
Using the result of Goldner et al \cite{GoldnerAhlers1992} for $\zeta$ we obtain
$3\zeta + \alpha -2 = -0.0012 \pm 0.0019$ where the
errors have been combined in quadrature. The agreement with
the prediction is very good.
Somewhat worse agreement is obtained if we use the result
of Adriaans \cite{AdriaansLipa1994} which leads to
$3\zeta + \alpha -2 = -0.0022 \pm 0.0005$. We note that
in both these experiments, the plots of deviations of the
measurements from the fits show systematic departures, indicating
that the error bars quoted may be
optimistic. We also note that our results are in good agreement with the
expected behavior based on vortex-ring RG theory. \cite{Williams1995}

\subsection{\label{sec:level6d}Thermal Conductivity}
A secondary goal of the experiment was to obtain thermal
conductivity results in the normal phase of helium above T$_{\lambda}$.
As described in section VI-B, a series of time
constants is derived from the fit to the thermal relaxation
transient after a heat pulse. The thermal
conductivity is inversely proportional to $\tau_{1}$ and can
therefore be output as part of the heat pulse
analysis.  As expected, we observe that the time constant
of the decay is shortest near
T$_{\lambda}$ ($\sim$ few seconds) and grows as
the temperature rises. Very near T$_{\lambda}$ where the
pulse size is small, the transient is difficult to
analyze: the small amplitude and short time constant are masked
by the HRT noise, especially by the larger cosmic ray spikes
which look very similar. This limited our analysis of
$\tau_{1}$ to $|t| > 10^{-9}$. The results for the thermal
conductivity are shown in Fig. 19 after being binned at
a density of 5 bins per decade of $t$. For comparison
we show some results obtained in a ground experiment that 
covers part of the region. \cite{DerivedfromLipaCzech1996} Also
shown is the behavior predicted by 
Dohm \cite{DohmPRevB1991} using RG techniques.
It can be seen that there is
good agreement between the two sets of data and
the theoretical model. \cite{theorynoteofref8}

%****************************************************************
\begin{figure}
\includegraphics{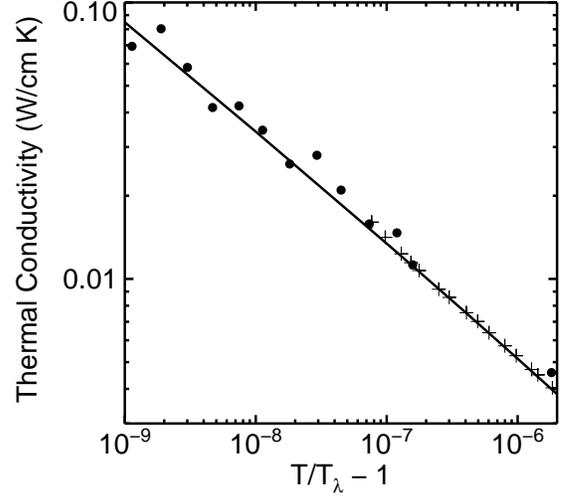}
\caption{\label{fig:epsart19}Log-log plot of thermal conductivity vs reduced temperature
above the lambda point. Filled circles: present work; $+$ :Lipa and Li; \cite{DerivedfromLipaCzech1996}
curve: model of Dohm. \cite{DohmPRevB1991}}
\end{figure}
%****************************************************************

%****************************************************************
\begin{figure}
\includegraphics{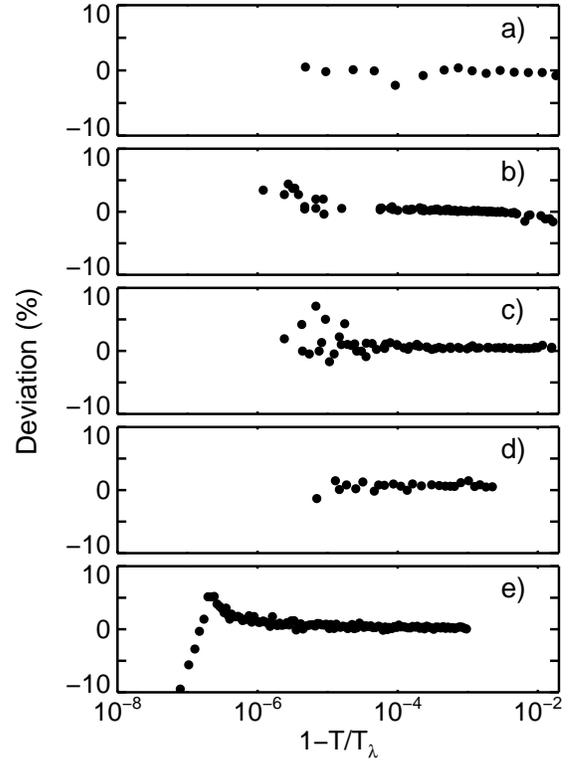}
\caption{\label{fig:epsart20}Deviations of various data sets from the best fit function vs
reduced temperature on a logarithmic scale, below the transition.
a: Buckingham, Fairbank and Kellers; \cite{FairbankKellers1957}
b: Ahlers; \cite{AhlersPRevA1971}
c: Gasparini and Moldover; \cite{GaspariniMoldover1975}
d: Takada and Watanabe \cite{TakadaJLowTempP1980}
 and e: Lipa and Chui. \cite{LipaPhyRevLett1983} }
\end{figure}
%****************************************************************

\subsection{\label{sec:level6e}Comparison with Other Measurements}
In Fig. 20 we show the deviations of other sets of data from our best fit
to Eq.(9). 
The data of 
Lipa and Chui \cite{LipaPhyRevLett1983} were obtained with a 3 mm
high cell and that of 
Ahlers \cite{AhlersPRevA1971} with a 1.59 cm high cell. 
Although it is
possible to correct for the effects of gravity, it can be seen that in the ground
experiments the corrections become quite large near the 
transition, limiting access to the transition region.
In Fig. 20(a,c,d) we show the deviations of other reported
results \cite{GaspariniMoldover1975,FairbankKellers1957,TakadaJLowTempP1980} 
from the same function. It can be
seen that there is reasonably good agreement between the various data sets, 
but occasionally the deviations are 
outside the noise of the data. Some of the discrepancies may be due to the use of earlier
temperature scales, but it is unlikely that this effect contributes more 
than $0.2\%$ to the deviations.

The significant improvement in scatter of the new data for $t > 10^{-7}$ can be attributed to
the use of the HRTs and a larger sample ($\sim 0.8$ moles). 
The larger sample size does not directly
improve the signal to noise of a heat capacity measurement, instead it attenuates the variations in the 
external heat leaks, leading to more accurate extrapolations to the 
center of a pulse.

\section{\label{sec:level7}CONCLUSION \protect\\}

In summary, our primary result is a new value of the 
exponent $\alpha$ describing the 
divergence of the specific heat below the lambda-transition of
helium with a reduced systematic uncertainty. 
This value is in good agreement with estimates
based on RG calculations and is also in reasonable
agreement with the value expected from superfluid density data via the Josephson scaling
relation. Our result for the ratio A$^{+}$/A$^{-}$ is also in good
agreement with recent estimates.
Thermal conductivity measurements derived from the
relaxation data are in reasonable agreement with steady state
measurements further from the transition 
and with theoretical predictions.

This gives us additional confidence in the current model
of second order transitions, and in the use
of the results for other analyses.
Other aspects of the RG theory, predicting the properties
of condensed matter in general, need to be
examined on their merits, but it is clear that the underlying principles are capable of 
explaining a wide range of properties of ordinary matter.
We note that the reduction of the thermometer noise caused by
cosmic rays that was demonstrated on a more recent 
mission \cite{LipaPhyRevLett2000} coupled with continuing advances
in thermometry \cite{KlemmeJLowTempP1999} opens up
the possibility of further substantial reductions
in the uncertainty of $\alpha$ in future experiments.

\section{\label{sec:level8}ACKNOWLEDGEMENTS \protect\\}

We wish to thank the NASA Office of Biological and Physical Research
for its generous support with contract $\#$JPL - 957448, and Dr. M. Lee for
his long term support of the program.  We greatly appreciate the support of the LPE teams at Stanford
and JPL, and the services of the KSC and MSFC mission support teams. We also thank Ball Aerospace for
extensive support with the development of the flight hardware, and especially G. Mills for
the fabrication of the calorimeter. Lastly, we thank the crew of the Space Shuttle Columbia on
STS-52 for their support during the microgravity periods of the mission.


\begin{thebibliography}{99}

%1
\bibitem{Wilson1971}K. G. Wilson, Phys. Rev. B  \textbf{4}, 3174 (1971).
For a recent
textbook on the subject, see H. Kleinert and V. Schulte-Frohlinde,
\textit{Critical Properties
of $\phi^{4}$ Theories} (World Scientific, Singapore, 2001).
%2
\bibitem{Moldover1979}M. R. Moldover, J. V. Sengers, R. W. Gammon and R. J. Hocken, Rev.
Mod. Phys., \textbf{51}, 79 (1979).
%3
\bibitem{HauptStraub1999}A. Haupt and J. Straub, Phys. Rev. E, \textbf{59}, 1795 (1999).
%4
\bibitem{Edwardsthesis1984}T. J. Edwards, thesis, University of
Western Australia, 1984 (unpublished). See also Table I of ref. 3.
%5
\bibitem{LipaPhyRevLett1983}J. A. Lipa and T. C. P. Chui, Phys. Rev. Lett., \textbf{51}, 2291 (1983).
%6
\bibitem{Lipa75thJubilee1983}J. A. Lipa, \textit{75th Jubilee Conference on Helium-4}, ed: J. G. M.
Armitage (World Scientific, Singapore, 1983), p. 208.
%7
\bibitem{Kleinert2000}H. Kleinert, Physics Letts. A, \textbf{277}, 205 (2000).
 See also ch.  19 in the
textbook cited in ref. 1.
%8
\bibitem{Campostrini2001}M. Campostrini, M. Hasenbusch, A. Pelissetto, P. Rossi and E.
Vicari, Phys. Rev. B, \textbf{63}, 214503 (2001).
%9
\bibitem{StrosserDohm2000}See ref. 47 in M. Str\"osser and V. Dohm, Phys. Rev. E, \textbf{67},
056115 (2003).  See also the three-loop results in 
M. Str\"osser, M. M\"onnigmann and V. Dohm, Physica B,
\textbf{284-288}, 41 (2000) and H. Kleinert and B. Van den Bossche, Phys. Rev. E \textbf{63},
056113 (2001).
%10
\bibitem{LipaPhyRevLett1996}J. A. Lipa, D. R. Swanson, J. A. Nissen, T. C. P. Chui and U. E.
Israelsson, Phys. Rev. Lett., \textbf{76}, 944 (1996).
%11
\bibitem{LipaPhyRevLett2000}See also footnote 15 of J. A. Lipa, D. R. Swanson, J. A. Nissen,
Z. K. Geng, P. R. Williamson, D. A. Stricker, T. C. P. Chui,  U. E. Israelsson and M. Larson,
Phys. Rev. Lett., \textbf{84}, 4894 (2000).
%12
\bibitem{Wegner1972}F. Wegner, Phys. Rev. B, \textbf{5}, 4529 (1972).
%13
\bibitem{Privam1991}V. Privman, P. C. Hohenberg and A. Aharony, 
in \textit{Phase Transitions and Critical Phenomena}, 
eds: C. Domb and J. L. Lebowitz (Acad. N.Y. 1991), \textbf{14}, p. 1.
%14
\bibitem{BarmatzPRevB1975}M. Barmatz, P.C. Hohenberg and A. Kornblit, 
Phys. Rev. B, \textbf{12}, 1947 (1975).
%15
\bibitem{MuellerPhyRevB1976}K. H. Mueller, G. Ahlers and F. Pobell, Phys. Rev. B, \textbf{14}, 2096 (1976).
%16
\bibitem{GaspariniMoldover1975}F. M. Gasparini and M. R. Moldover, Phys. Rev.  B,  \textbf{12}, 93 (1975).
%17
\bibitem{GaspariniGaeta1978}F. M. Gasparini and A. A. Gaeta, Phys. Rev.  B,  \textbf{17}, 1466 (1978).
%18
\bibitem{GoldnerAhlers1992}L. S. Goldner, N. Mulders and G. Ahlers, J. Low Temp. Phys., \textbf{93}, 131 (1992).
%19
\bibitem{AdriaansLipa1994}M. J. Adriaans, D. R. Swanson and J. A. Lipa, Physica B, \textbf{194-196}, 733 (1994).
%20
\bibitem{Ahlersbook1976}G. Ahlers in: \textit{The Physics of Liquid and Solid Helium}, eds: K. H.
Bennemann and J. B. Ketterson (Wiley, N.Y. 1976) Part \textbf{1}, p. 85.
%21
\bibitem{SchlomsDohmEuroLett1987}R. Schloms and V. Dohm, Europhys. Letts., \textbf{3}, 413 (1987).
%22
\bibitem{GreywallAhlers1973}D. S. Greywall and G. Ahlers, Phys. Rev. A, \textbf{7}, 2145 (1973).
%23
\bibitem{SingsaasAhlers1984}A. Singsaas and G. Ahlers, Phys. Rev. B, \textbf{30}, 5103 (1984).
%24
\bibitem{SchlomsDohm1990}R. Schloms and V. Dohm, Phys. Rev. B, \textbf{42}, 6142 (1990).
%25
\bibitem{GuidaJPhysA1998}R. Guida and J. Zinn-Justin, J. Phys. A, \textbf{31}, 8103 (1998).
%26
\bibitem{FairbankKellers1957}W. M. Fairbank, M. J. Buckingham and C. F. Kellers, \textit{Proc. 5th Int.
Conf. on Low Temp. Phys}. (Madison, Wis. 1957), p. 50.
%27
\bibitem{Ahlers1968Barmatz1968}See for example, G. Ahlers, Phys. Rev., \textbf{171}, 275 (1968); and M.
Barmatz and I. Rudnick, Phys. Rev., \textbf{170}, 224 (1968).
%28
\bibitem{LipaCryogen1994}J. A. Lipa, D. R. Swanson, J. A. Nissen and T. C. P. Chui,
Cryogenics, \textbf{34}, 341 (1994).
%29
\bibitem{LuchikElliottCryoEng1996}T.S. Luchik, U. E. Israelsson, D. Petrac and S. Elliott, Adv. Cryo.
Eng., \textbf{41}, 1135 (1996).
%30
\bibitem{NissenLipaAIAA}J. A. Nissen, D. R. Swanson, D. A. Stricker and J. A. Lipa, AIAA paper $\#2000-0943$.
%31
\bibitem{Nippon6N}Nippon Mining Co. Japan, grade 6N (99.9999$\%$ pure).
%32
\bibitem{HurstLankford1984}J. G. Hurst and A. B. Lankford, U.S. Department of Commerce NBSIR 84-3007,
(1984).
 See also S. S. Rosenblum, W. A. Steyert and F. R. Fickett, Cryogenics, \textbf{17}, 645 (1977).
%33
\bibitem{LipaPhysica1981Chui1992ChuiPRevLett1992}J. A. Lipa, B. C. Leslie 
and T. C. Wallstrom, Physica  \textbf{107} B, 331
(1981);  T. C. P. Chui, D. R. Swanson, M. J. Adriaans, J. A. Nissen, and J. A. Lipa, 
\textit{Temperature, its Measurement and Control in Science and Industry},
\textbf{6}, 1213 (1992) and T. C. P. Chui, D. R. Swanson, M. J. Adriaans, J. A. Nissen, 
and J. A. Lipa, Phys. Rev. Lett.,  \textbf{69}, 3005 (1992).
%34
\bibitem{Velu1976}E. Velu, J. -P. Renard and B. Lecuyer, Phys. Rev. B, \textbf{14}, 5088 (1976).
%35
\bibitem{SwansonLipa1994}D. R. Swanson, J. A. Nissen, T. C. P. Chui, P. R. Williamson and J.
A. Lipa, Physica B, \textbf{194}, 25 (1994).
%36
\bibitem{MarekJap1987}D. Marek, Jap. J. App. Phys., \textbf{26}, 1683 (1987).
%37
\bibitem{WilliamsCorp}Williams Mfg. Corp., San Jose, Calif.
%38
\bibitem{LT17LuuLipa1984}J. Luu, T. C. P. Chui and J. A. Lipa,  \textit{Proc. Internat. Conf. on Low
Temp. Phys.(LT-17)}, eds: U.Eckern et al. (Karlsruhe, 1984), p. 207.
%39
\bibitem{RigbyChui1990}K. W. Rigby, D. Marek and T. C. P. Chui, Rev. Sci. Instrum., \textbf{61},
834 (1990).
%40
\bibitem{Germany}Vacuumschmelze GMBH, Germany.
%41
\bibitem{BiomagTech}Biomagnetic Technologies Inc., San Diego, CA.
%42
\bibitem{MasonAdvCryoEng1980}P. Mason, D. Collins, P. Cowgill, D. Elleman, D. Petrac,
M. Saffren and T. Wang, Adv. Cryo. Eng., \textbf{25}, 801 (1980).
%43
\bibitem{PetracLuchik1994}D. Petrac, U. E. Israelsson, and T. S. Luchik, Adv. Cryo. Eng., \textbf{39},
137 (1994).
%44
\bibitem{RogersAdvSpace1998}See, for example, M. J. B. Rogers, K. Hrovat and M. Moskowitz, 
Adv. Space Res., \textbf{22}, No. 8, 1257 (1998).
%45
\bibitem{LakeshoreModelSN}Lakeshore Cryotronics, Westerville, OH, Model $\#$ RF-800-2, SN. B109B.
%46
\bibitem{LakeshoreKrange}Lakeshore Cryotronics, 1.2 - 22 K range.
%47
\bibitem{PrestonMetro1990}H. Preston-Thomas, Metrologia, \textbf{27}, 3 (1990), and ibid. p. 107.
%48
\bibitem{RubinSciInst1972}L. G. Rubin and Y. Golahny, Rev. Sci. Instrum., \textbf{43}, 1758 (1972).
%49
\bibitem{ElectroSciIndModelSN}Electro Scientific Industries, Portland, OR, Model $\#$ RS-925, SN.
623012.
%50
\bibitem{LeungCryog1979}Y. K. Leung and J. F. Kos, Cryogenics, \textbf{19}, 531 (1979).
%51
\bibitem{CallenPhysRev1951}H. B. Callen and T. A. Welton, Phys. Rev., \textbf{83}, 34 (1951).
%52
\bibitem{QinChuiCryogen1996}X. Qin, J. A. Nissen, D. R. Swanson, P. R. Williamson, D. A.
Stricker, J. A. Lipa,  T. C. P. Chui
and U.E. Israelsson, Cryogenics, \textbf{36}, 781 (1996).
%53
\bibitem{KerrAnnPhys1964}E. C. Kerr and R. D. Taylor, Ann. Phys., \textbf{26}, 292 (1964);
J. J. Niemela and R. J. Donnelly, J. Low Temp. Phys., \textbf{98}, 1 (1995).
%54
\bibitem{KellersDuke1960}C. F. Kellers, thesis, Physics Dept., Duke University, Durham, NC,
(1960).
%55
\bibitem{AhlersPRevA1971}G. Ahlers, Phys. Rev. A, \textbf{3}, 696 (1971). (We note some typographical
errors in Eq's (7) and (10)
 of this paper.)
%56
\bibitem{NissenLipaCzech1996}J. A. Nissen, D. R. Swanson, X. Qin and J. A. Lipa,  Czech. J.
Phys., \textbf{46-S1},  379 (1996).
%57
\bibitem{WilliamsonLipa1995Cosmic}P. R. Williamson, J. A. Nissen,
D. R. Swanson and J. A. Lipa, \textit{Proc.
24th Int. Cosmic Ray Conf}.
(Rome, Italy, Aug. '95), \textbf{4}, p.1287.
%58
\bibitem{NumericalRecipes1987}W. H. Press, B. P. Flannery, S. A. Teukolsky and W. T. Vetterlling,
\textit{Numerical Recipes}
(Cambridge University Press, Cambridge, 1987), p. 539.
%59
\bibitem{OzisikBVPHeatCond1989}M. Necati Ozisik, \textit{Boundary Value Problems of Heat Conduction},
(Dover, NY 1989).
%60
\bibitem{epapsdata}The complete set of raw data used in the curve fitting is listed in
EPAPS Document No. [number to be inserted by publisher]. Also listed is the
bin-averaged data set shown in Fig 15. A direct link to this document may be found
in the online article's HTML reference section. The document may also be reached via 
the EPAPS homepage (http://www.aip.org/pubservs/epaps.html) 
or from ftp.aip.org in the directory /epaps/. See the EPAPS homepage for more information. 
%61
\bibitem{Bevington}P. R. Bevington, \textit{Data Reduction and Error Analysis for the 
Physical Sciences}, (McGraw-Hill Book Co., NY, 1969), p. 66.
%62
\bibitem{KleinertPRevD1999}H. Kleinert, Phys. Rev. D, \textbf{60}, 085001 (1999).
%63
\bibitem{StrosserDohmPRevE2003}M. Str\"osser and V. Dohm, Phys. Rev. E (to be published).
%64
\bibitem{AlhersComm}G. Ahlers, private communication.
%65
\bibitem{Williams1995}G. A. Williams, J. Low Temp. Phys., \textbf{101}, 421 (1995).
%66
\bibitem{DerivedfromLipaCzech1996}Derived from the results in: J. A. Lipa and Q. Li, Czech. J. Phys.,
\textbf{46}, 185 (1996) Suppl. S1.
%67
\bibitem{DohmPRevB1991}V. Dohm, Phys. Rev. B, \textbf{44}, 2697 (1991).
%68
\bibitem{theorynoteofref8}We note that the theoretical
curve in Fig. 3 of ref. 10 was drawn
slightly too low.
%69
\bibitem{TakadaJLowTempP1980}T. Takada and T. Watanabe, 
J. Low Temp. Phys., \textbf{41}, 221 (1980).
%70
\bibitem{KlemmeJLowTempP1999}B. J. Klemme, M. J. Adriaans, P. K. Day, D. A. Sergatskov, T. L.
Aselage and R. V. Duncan, J.
 Low Temp. Phys., \textbf{116}, 133 (1999).

\end{thebibliography}
\end{document}